\documentclass[10pt,letterpaper]{article}
\usepackage[top=0.85in,left=2.75in,footskip=0.75in]{geometry}

\usepackage{amsmath,amssymb}

\usepackage{changepage}

\usepackage[utf8x]{inputenc}

\usepackage{textcomp,marvosym}

\usepackage{cite}

\usepackage{nameref,hyperref}

\usepackage[right]{lineno}

\usepackage{microtype}
\DisableLigatures[f]{encoding = *, family = * }

\usepackage[table]{xcolor}

\usepackage{array}

\newcolumntype{+}{!{\vrule width 2pt}}

\newlength\savedwidth

\raggedright
\usepackage[aboveskip=1pt,labelfont=bf,labelsep=period,justification=raggedright,singlelinecheck=off]{caption}

\makeatletter
\renewcommand{\@biblabel}[1]{\quad#1.}
\makeatother

\usepackage{lastpage,fancyhdr,graphicx}
\usepackage{epstopdf}
\fancyheadoffset[L]{2.25in}
\fancyfootoffset[L]{2.25in}
\usepackage{nicefrac} % pretty inline fractions
\def\t{{^\text{T}}} % transpose
\newcommand{\KBF}{} % Kevin comments
\newcommand{\JHL}{} % John L. comments
\definecolor{lightblue}{HTML}{ADD8E6} % for cell colors
\usepackage{soul} % for striking out text \st{}

\begin{document}
\vspace*{0.2in}

% Title must be 250 characters or less.
\begin{flushleft}
{\Large
\textbf\newline{Biologically-informed neural networks guide mechanistic modeling from sparse experimental data} % Please use "sentence case" for title and headings (capitalize only the first word in a title (or heading), the first word in a subtitle (or subheading), and any proper nouns).
}
\newline
% Insert author names, affiliations and corresponding author email (do not include titles, positions, or degrees).
\\
John H. Lagergren\textsuperscript{1,2*},
John T. Nardini\textsuperscript{1,3},
Ruth E. Baker\textsuperscript{4},
Matthew J. Simpson\textsuperscript{5},
Kevin B. Flores\textsuperscript{1,2*}
\\
\bigskip
\textbf{1} Department of Mathematics, North Carolina State University, Raleigh, North Carolina, USA
\\
\textbf{2} Center for Research and Scientific Computation, North Carolina State University, Raleigh, North Carolina, USA
\\
\textbf{3} Statistical and Applied Mathematical Sciences Institute, Durham, North Carolina, USA
\\
\textbf{4} Mathematical Institute, University of Oxford, Oxford OX2 6GG, UK
\\
\textbf{5} School of Mathematical Sciences, Queensland University of Technology, Brisbane, Queensland 4001, Australia
\\
\bigskip

% Insert additional author notes using the symbols described below. Insert symbol callouts after author names as necessary.
% 
% Remove or comment out the author notes below if they aren't used.
%
% Primary Equal Contribution Note
%\Yinyang These authors contributed equally to this work.

% Additional Equal Contribution Note
% Also use this double-dagger symbol for special authorship notes, such as senior authorship.
%\ddag These authors also contributed equally to this work.

% Current address notes
%\textcurrency Current Address: Dept/Program/Center, Institution Name, City, State, Country % change symbol to "\textcurrency a" if more than one current address note
% \textcurrency b Insert second current address 
% \textcurrency c Insert third current address

% Deceased author note
%\dag Deceased

% Group/Consortium Author Note
%\textpilcrow Membership list can be found in the Acknowledgments section.

% Use the asterisk to denote corresponding authorship and provide email address in note below.
*jhlagerg@ncsu.edu, kbflores@ncsu.edu

\end{flushleft}

% temporary
%\tableofcontents
%\newpage

% Please keep the abstract below 300 words
\section*{Abstract}
Biologically-informed neural networks (BINNs), an extension of physics-informed neural networks~\cite{raissi_2019}, are introduced and used to discover the underlying dynamics of biological systems from sparse experimental data. In the present work, BINNs are trained in a supervised learning framework to approximate \textit{in vitro} cell biology assay experiments while respecting a generalized form of the governing reaction-diffusion partial differential equation (PDE). By allowing the diffusion and reaction terms to be multilayer perceptrons (MLPs), the nonlinear forms of these terms can be learned while simultaneously converging to the solution of the governing PDE. Further, the trained MLPs are used to guide the selection of biologically interpretable mechanistic forms of the PDE terms which provides new insights into the biological and physical mechanisms that govern the dynamics of the observed system. The method is evaluated on sparse real-world data from wound healing assays with varying initial cell densities~\cite{simpson_2016}.

% Please keep the Author Summary between 150 and 200 words
% Use first person. PLOS ONE authors please skip this step. 
% Author Summary not valid for PLOS ONE submissions.   
\section*{Author summary}
In this work we extend equation learning methods to be feasible for biological applications with nonlinear dynamics and where data are often sparse and noisy. Physics-informed neural networks have recently been shown to approximate solutions of PDEs from simulated noisy data while simultaneously optimizing the PDE parameters. However, the success of this method requires the correct specification of the governing PDE, which may not be known in practice. Here, we present an extension of the algorithm that allows neural networks to learn the nonlinear terms of the governing system without the need to specify the mechanistic form of the PDE. Our method is demonstrated on real-world biological data from scratch assay experiments and used to discover a previously unconsidered biological mechanism that describes delayed population response to the scratch. 

%\linenumbers

% Use "Eq" instead of "Equation" for equation citations.
% For figure citations, please use "Fig" instead of "Figure".
% Place figure captions after the first paragraph in which they are cited.
% Results and Discussion can be combined.
% Place tables after the first paragraph in which they are cited.
%PLOS does not support heading levels beyond the 3rd (no 4th level headings).

\section*{Introduction}

%The Introduction should put the focus of the manuscript into a broader context. As you compose the Introduction, think of readers who are not experts in this field. Include a brief review of the key literature. If there are relevant controversies or disagreements in the field, they should be mentioned so that a non-expert reader can delve into these issues further. The Introduction should conclude with a brief statement of the overall aim of the experiments and a comment about whether that aim was achieved.

%
% Math modeling of collective migration
%

\emph{Collective migration} refers to the coordinated migration of a group of individuals~\cite{friedl_collective_2009, vicsek_novel_1995}. This process arises in a variety of biological and social contexts, including pedestrian dynamics~\cite{helbing_social_1995}, tumor progression~\cite{gallaher_evolution_2013}, and animal development~\cite{mclennan_neural_2015}. In the presence of many individuals, differential equation models provide a flexible framework to investigate collective behavior as a continuum~\cite{arciero_continuum_2011, dyson_importance_2015, johnston_mean-field_2012, nardini_modeling_2016, topaz_locust_2012}. A challenge for mathematicians and scientists is to use mathematical models together with spatiotemporal data of collective migration to validate assumptions about the underlying physical and biological laws that govern the observed dynamics. \JHL{Several factors contribute to the difficulty of this task, even for simple systems/data, some of which include biological forms and levels of noise in the observation process, poor understanding of the underlying dynamics, a large number of candidate mathematical models, implementation of computationally expensive numerical solvers, etc. This work provides a data-driven tool which can alleviate many of these problems by enabling the rapid development and validation of mathematical models from sparse noisy data. The methodology is demonstrated using a case study of scratch assay experiments.}

Scratch assays are a widely adopted experiment in cellular biology used to study collective cell migration \textit{in vitro} as cell populations re-colonize empty spatial regions. These experiments have been used previously to observe population-wide behavior in many different contexts, including wound healing~\cite{aoki_propagating_2017, chapnick_leader_2014, matsubayashi_erk_2004, nikolic_role_2006} and cancer progression~\cite{haridas_quantifying_2017}. Mathematical modeling of scratch assays plays an important role in the quantification and analysis of population dynamics. This is because (i) the equations and parameters comprising mathematical models are \emph{interpretable}, providing information about the underlying physical and biological mechanics that drive the observed system, and (ii) when properly calibrated, they are \emph{generalizable}, affording the ability to make accurate predictions beyond the data set used for calibration. 

Reaction-diffusion partial differential equations (PDEs) are frequently used to model scratch assay experiments~\cite{simpson_2016, arciero_continuum_2011, nardini_modeling_2016, johnston_estimating_2015, maini_travelling_2004}. The general one-dimensional reaction-diffusion equation that describes the rate of change of a quantity of interest $u(x,t)$ (e.g. cell density) is 
\begin{eqnarray}
\label{eq:reaction_diffusion}
    u_t = (\mathcal{D}u_x)_x + \mathcal{G}u, \quad x\in[x_0,x_f], \quad t\in[t_0,t_f],
\end{eqnarray}
in which the rate of change of $u$ (i.e. $u_t$) is a function of diffusion, modeled by the function $\mathcal{D}$, and reaction or growth, modeled by the function $\mathcal{G}$. Note that $\mathcal{D}$ and $\mathcal{G}$ \JHL{depend on the application, and choosing the correct/optimal mechanistic models for these terms is the focus of many current research efforts and remains an open question.} The classical Fisher–Kolmogorov–Petrovsky–Piskunov (FKPP) equation is a reaction-diffusion equation that has been used to model a wide spectrum of growth and transport of biological processes. In particular, the FKPP model assumes a scalar diffusivity function $\mathcal{D} = D$ and logistic growth function $\mathcal{G} = r(1-u/K)$ with intrinsic growth rate $r$ and carrying capacity $K$~\cite{baldock_patient-specific_2014, maini_travelling_2004}. Variants of the reaction-diffusion equation have also been used to account for different types of cell interactions during scratch assay experiments. For example, the nonlinear diffusivity function $\mathcal{D} = 1 - \alpha\nicefrac{4}{3} + 3\alpha\left(u/K - \nicefrac{2}{3}\right)^2$ with cell-to-cell adhesion coefficient $\alpha$ was used to model dynamics in which neighboring cells prevent other cells from migrating~\cite{anguige_one-dimensional_2009}. Alternatively, a diffusivity function of the form $\mathcal{D} = D(1+\alpha(u/K)^2)$ can be used to model dynamics in which cells promote the migration of others~\cite{nardini_modeling_2016}. Additional variants of reaction-diffusion equation models have captured cell migration in the presence of growth factors~\cite{dale_travelling_1995}, during melanoma progression~\cite{haridas_quantifying_2017}, and in response to different drug treatments~\cite{johnston_estimating_2015}.

A recent study quantitatively investigated the role of initial cell density by conducting a suite of scratch assay experiments on PC-3 prostate cancer cells with systematically varying initial cell densities~\cite{simpson_2016}. The experimental data was used to calibrate the FKPP equation as well as a variant model known as the Generalized Porous-FKPP equation, which assumes that diffusivity increases with cell density $u$ by using a diffusivity function $\mathcal{D} = D(u/K)^m$ with diffusion coefficient $D$, carrying capacity $K$, and exponent $m$. Like the FKPP equation, the growth term is also described by the logistic growth function $\mathcal{G} = r(1-u/K)$. While the calibrated models approximated the experimental data well in many cases in~\cite{simpson_2016}, the presence of systematic biases between the model solutions and experimental data indicate the existence of additional governing mechanisms that may not be accounted for in these mathematical models. However, the existence of a large number of possible biophysical mechanisms that could play a role in scratch assay dynamics makes the testing of mathematical models against these experimental data computationally challenging. Thereby, this scenario motivates the use of equation learning methods to discover the diffusion and reaction terms directly from the experimental data. 

%
% Equation learning methods
%

Enabled by advances in computing power, algorithms, and the amount of available data, the field of equation learning has recently emerged as a powerful tool for the automated identification of underlying physical laws governing a set of observation data. The basic assumption in this field is that measured data arise from some unknown $n$-dimensional dynamical system of the form
\begin{eqnarray}
\label{eq:governing_dynamics}
    u_t = \mathcal{F}(x, t, u, u_{x}, u_{xx}, \dots; \theta), \quad x\in[x_0,x_f], \quad t\in[t_0,t_f],
\end{eqnarray}
with quantity of interest $u=u(x, t)$, parameter vector $\theta\in\mathbb{R}^k$, and appropriate initial and boundary conditions. \JHL{An example quantity of interest for modeling cell migration dynamics is the cell density ($\nicefrac{\text{cells}}{\text{mm}^2}$) at location $x$ and time $t$.} The measured data $\{u_{i,j}\}_{i,j=1}^{M,N}$ for a set of spatial points $x_i$, $i=1,\dots,M$, and set of time points $t_j$, $j=1,\dots,N$, are assumed to be corrupted by some form of observation error that may be known or unknown in practice. The goal of equation learning methods is to identify the closed form of $\mathcal{F}$ in Eq~\eqref{eq:governing_dynamics} directly from the noisy measurements $u_{i,j}$. \JHL{Note that, in order to simulate the learned equation, either the noisy or a denoised version of the initial condition can be used along with an assumed boundary condition (e.g. no-flux) that describes the biological process generating the data.}

Two primary sets of methodology have been used in field of equation learning to date: sparse regression~\cite{brunton_2016, rudy_2017} and theory-informed neural networks~\cite{raissi_2019, yang_2020}. In the sparse regression framework, numerical methods (e.g. finite differences or polynomial splines) are used to denoise $u$ and approximate the partial derivatives $u_t$, $u_x$, $u_{xx}$, etc. from a set of data. The approximations are then used to construct a library of nonlinear candidate terms (e.g. 1, $u$, $u^2$, $u_x$, $\dots$, $u_x^2u_{xx}^2$, etc.) thought to comprise the governing system of ordinary differential equations (ODEs) or PDEs. The data relating $u_t$ to all possible model terms inside the library are formulated as a linear regression problem in which sparsity promoting techniques are used to select a small subset of library terms that produce the most parsimonious model. While the sparse regression framework has been successfully demonstrated to circumvent searching through a combinatorially large space of possible candidate models, it can require large amounts of training data and the numerical methods used for denoising and differentiation are not robust to biologically realistic forms and levels of noise, leading to inaccuracies in both the constructed library and learned equations~\cite{lagergren_2020}. \JHL{Further, the method assumes the unknown function $\mathcal{F}$ in Eq~\eqref{eq:governing_dynamics} can be written as a linear combination of nonlinear candidate terms, which may not be true in practice.}

An alternative approach uses function-approximating deep neural networks, i.e., multilayer perceptrons (MLPs), as surrogate models $u_{\text{MLP}}(x, t)$ for the solution of the governing dynamical system~\cite{raissi_2019, yang_2020}. In this approach, the assumed mechanistic form of $\mathcal{F}$ in Eq~\eqref{eq:governing_dynamics} is pre-specified and then used as a form of regularization in the neural network objective function. The parameters of $\mathcal{F}$ are allowed to be ``learnable,'' meaning that the parameters of the governing PDE are calibrated while the neural network is trained to minimize the error between $u_{\text{MLP}}(x_i, t_j)$ and the data $u_{i, j}$. This methodology ensures that the neural network solution satisfies the physical laws described by $\mathcal{F}$ while simultaneously fitting the spatiotemporal data. Theory-informed neural networks have been demonstrated with smaller amounts of data in the presence of noise, however, they have so far only been applied to problems where the governing mechanistic PDE is known \textit{a priori}. 

Hybrid approaches that combine neural networks and sparse regression have also been suggested to address some of the issues surrounding the above methods~\cite{lagergren_2020, both_2019}. In these approaches, neural networks are used as surrogate models for $u(x,t)$ and then used to construct the library of candidate terms for sparse regression using automatic differentiation. These methods have been shown to accurately learn the governing system of equations for a variety of reaction-diffusion models from spatiotemporal data with biologically realistic levels of noise~\cite{lagergren_2020}. 

All three approaches (i.e. sparse regression, theory-informed neural networks, and hybrids) however, suffer from the \emph{model specification problem}, in which the governing ODE/PDE model must be specified \textit{a priori} either explicitly or as a library of candidate terms. Thus, (i) if the true dynamical system contains terms that are not included in the regularization term for theory-informed neural networks, or (ii) if the true terms cannot be represented as a linear combination of nonlinear candidate terms for sparse regression, then these methods will ultimately fail to recover the true system. \JHL{Further, detecting this issue when determining what the ``true'' system is in real-use cases is an open question.} Where systems with scalar or linear dynamics may be suitable for these approaches, biological systems pose a particular challenge in this respect, since many of the underlying mechanics driving these systems are nonlinear. For example, the Generalized Porous-FKPP model contains a nonlinear diffusivity function $\mathcal{D}=D(u/K)^m$) with unknown exponent $m$. These issues help explain why, to the best of our knowledge, equation learning methods have not yet been successfully applied to real-world biological population-level data. 

%
% Introducing BINNs to solve the model specification problem
%

In this work, biologically-informed neural networks (BINNs), an extension of physics-informed neural networks (PINNs)~\cite{raissi_2019}, are presented as a solution to the library specification problem for systems with biological/physical constraints. In this framework, the right-hand-side function $\mathcal{F}$ of the PDE in Eq~\eqref{eq:governing_dynamics} is assumed to be a combination of biologically relevant terms. For example, the general form of reaction-diffusion models can be described by the two right-hand-side terms in Eq~\eqref{eq:reaction_diffusion} meaning that the equation learning problem is transformed from learning $\mathcal{F}$ to learning the diffusivity and growth functions $\mathcal{D}$ and $\mathcal{G}$. Rather than assigning mechanistic forms to each function as in previous equation learning studies, each function is replaced with a separate neural network. This approach leverages the ability of deep neural networks to approximate continuous functions arbitrarily well~\cite{hornik_1991}. Importantly, the form of each learned neural network function can be visualized, thereby enabling a data-driven tool for user-guided conjecture of new mathematical equations that describe each separate term in $\mathcal{F}$. Moreover, formulating the equation learning task within the BINNs framework enables the modeler to use domain expertise to include qualitative constraints on the parameter networks (e.g. specifying nonlinear functions that are non-negative, monotone increasing/decreasing, etc.) by selecting appropriate activation functions and loss terms for the optimization. 

While BINNs can be used to discover a wide range of governing equations across the biological and physical sciences, including systems of ODEs and PDEs, in this work they are demonstrated using reaction-diffusion PDEs. The BINNs methodology is first tested using synthetic data and then demonstrated on experimental data from scratch assay experiments with variable initial cell densities~\cite{simpson_2016}. Notably, each data set is noisy and sparse, containing only five time measurements across 38 spatial locations. BINNs are used to discover the nonlinear forms of the diffusivity function and growth term of the governing reaction-diffusion equation. Persistent model discrepancy is used to motivate the incorporation of a novel delay term \JHL{which may have important implications for the reproducibility and modeling of scratch assays.} The learned nonlinear forms of the diffusion, growth, and delay terms are used to guide the selection of a mechanistic model with biologically interpretable parameters that remove virtually all of the model discrepancy. 

\subsection*{Scratch assay data}

\begin{figure}[!h]
    \centering
    \includegraphics[width=1.0\textwidth]{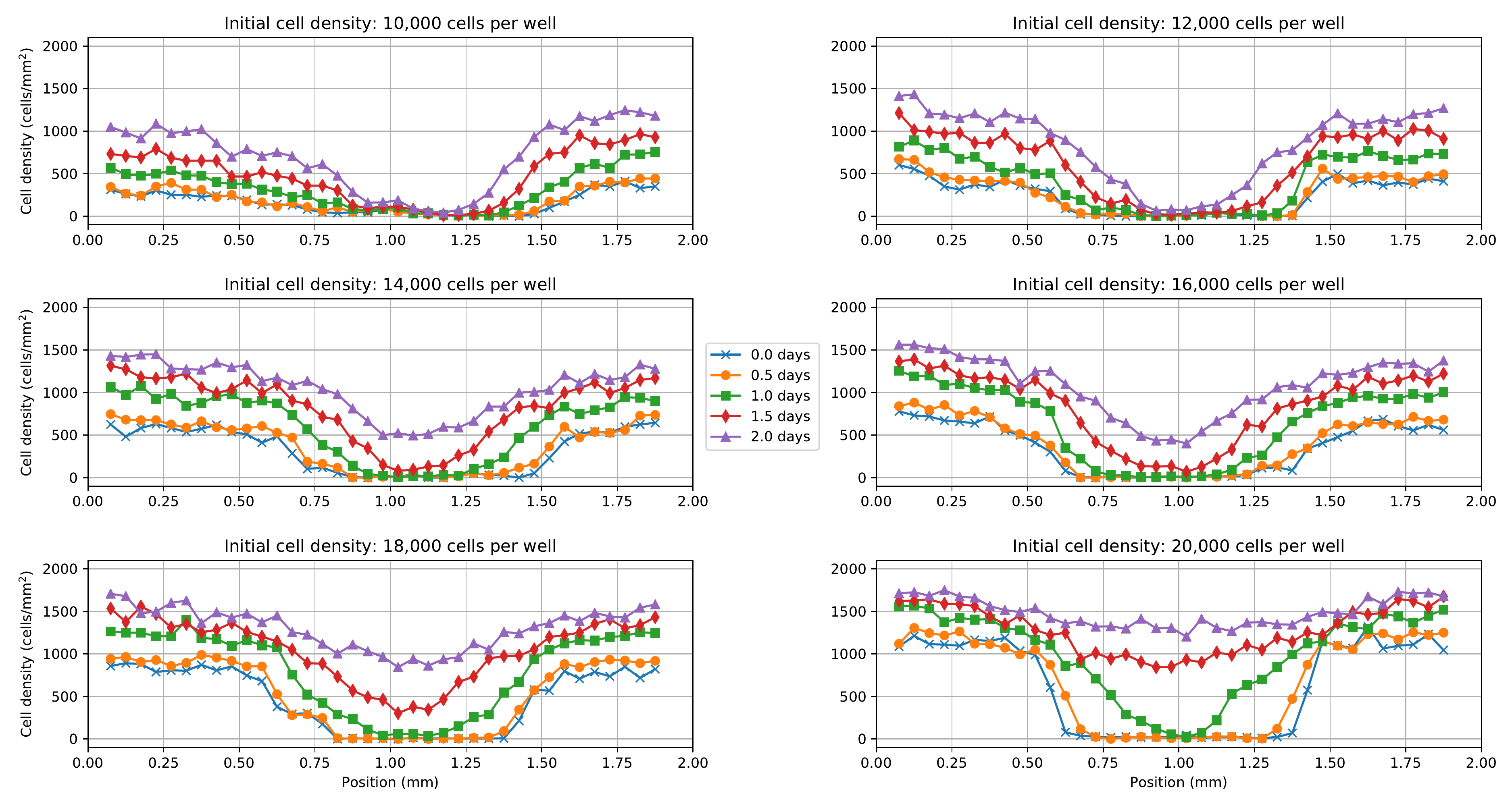}
    \caption{\textbf{Experimental scratch assay data.} Pre-processed cell density profiles from scratch assay experiments with varying initial cell densities~\cite{simpson_2016}. Each subplot corresponds to an experiment with a different initial cell density (i.e. 10,000, 12,000, 14,000, 16,000, 18,000, and 20,000 cells per well). The cell densities are reported at 37 equally-spaced positions and five equally-spaced time points.}
    \label{fig:raw_data}
\end{figure}

Biologically-informed neural networks (BINNs) are evaluated on experimental scratch assay data from~\cite{simpson_2016}. \KBF{A typical scratch assay involves (i) growing a cell monolayer up to some desired initial cell density, (ii) creating a ``scratch'' in the interior of the monolayer to produce an empty region, and (iii) recording longitudinal measurements of the cell density during re-colonization of the area. One-dimensional cell density profiles are obtained by manually counting the cells within vertical columns of the two-dimensional image data. See~\nameref{scratch_assay_experiment} for a visualization of the experiment.} For these data, the cell density profiles were reported for six varying initial cell density levels (i.e. 10,000, 12,000, 14,000, 16,000, 18,000, and 20,000 cells per well). To make the cell density profiles compatible with neural network training, the data are pre-processed by rescaling the $x$ and $t$ variables to the scales of millimeters (mm) and days, respectively (see Methods Section for more details). Further, the cell density profile at the left boundary is removed from the data because it was identified as an outlier across each of the six data sets. The resulting pre-processed cell densities at 37 spatial points and five time points are shown in Fig~\ref{fig:raw_data}.

\subsection*{Biologically-informed neural networks}\label{sec:binns}

\begin{figure}[!h]
    \centering
    \includegraphics[width=1.0\textwidth]{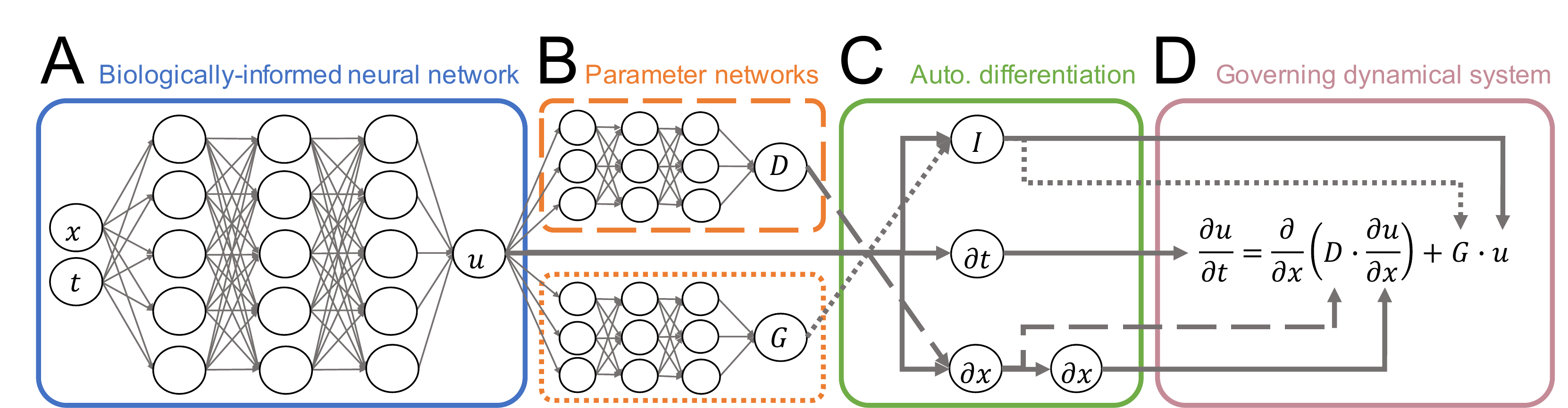}
    \caption{\textbf{Biologically-informed neural networks for reaction-diffusion models.} (A) BINNs are deep neural networks that approximate the solution of a governing dynamical system. (B) By allowing the terms of the dynamical system (e.g. diffusivity function $\mathcal{D}$ and growth function $\mathcal{G}$) to be function-approximating deep neural networks, the nonlinear forms of these terms can be learned without the need to specify a mechanistic model or library of candidate terms. (C) Automatic differentiation is used on compositions of the different neural network models (e.g. $u$, $D$, and $G$) to construct the PDE that describes governing dynamical system. (D) The governing system is used in the neural network objective function to jointly learn and satisfy the governing PDE while minimizing the error between the network outputs and noisy observations.}
    \label{fig:binn_schematic}
\end{figure}

BINNs are centered around a function-approximating deep neural network, or MLP, denoted by $u_{\text{MLP}}(x, t)$ which acts as a surrogate model that approximates the solution to the governing equation described by Eq~\eqref{eq:governing_dynamics} (Fig~\ref{fig:binn_schematic}A). In this work, the governing PDE is assumed to contain two terms, $\mathcal{D}$ and $\mathcal{G}$, that describe the general reaction-diffusion model in Eq~\eqref{eq:reaction_diffusion}. Since the true forms of the diffusivity and growth functions are unknown, they are approximated by neural networks $\mathcal{D} = D_{\text{MLP}}(u)$ and $\mathcal{G} = G_{\text{MLP}}(u)$ (Fig~\ref{fig:binn_schematic}B). Both $D_{\text{MLP}}$ and $G_{\text{MLP}}$ are continuously differentiable functions that input the predicted cell density $u_{\text{MLP}}(x,t)$ and output the corresponding diffusivity or growth value. The advantage of using MLPs for the terms of the governing PDE is that the nonlinear forms of these terms can be learned without specifying them explicitly (or as a library of candidate terms), thus circumventing the model specification problem. Automatic differentiation (Fig~\ref{fig:binn_schematic}C) is used to numerically differentiate compositions of $u_{\text{MLP}}$, $D_{\text{MLP}}$, and $G_{\text{MLP}}$ in order to construct the general reaction-diffusion model in Eq~\eqref{eq:reaction_diffusion}. The resulting PDE (Fig~\ref{fig:binn_schematic}D) is used to regularize $u_{\text{MLP}}$ during training so that $u_{\text{MLP}}$ not only fits the data $u_{i,j}$ but also satisfies the governing reaction-diffusion system.

To ensure that the fit to the data and the fidelity to the governing PDE are simultaneously optimized, the BINNs are trained with gradient-based methods using the following multi-part objective function:
\begin{eqnarray}
\label{eq:total_error}
    \mathcal{L}_{\text{Total}} = \mathcal{L}_{\text{GLS}} + \mathcal{L}_{\text{PDE}} + \mathcal{L}_{\text{Constr}}.
\end{eqnarray} 
%\noindent\textbf{GLS Loss.} 
The first term $\mathcal{L}_{\text{GLS}}$ concerns the generalized least squares (GLS) distance between $u_{\text{MLP}}(x_i, t_j)$ and the corresponding observed data $u_{i,j}$. The observation process is assumed to be described by a statistical error model of the form
\begin{equation}\label{eq:statistical_model}
    u_{i,j} = u(x_i, t_j) + w_{i,j}\odot\varepsilon_{i,j},
\end{equation}
in which the measured data $u_{i,j}$ are a combination of the underlying dynamical system $u(x_i, t_j)$ and some random variable $w_{i,j}\odot\varepsilon_{i,j}$ where $\odot$ represents element-wise multiplication~\cite{banks_2014}. In general, the independent and identically distributed (i.i.d.) random variable $\varepsilon_{i,j}$ is modeled by an $n$-dimensional normal distribution with mean zero and variance one that is weighted by 
\begin{eqnarray}
\label{eq:error_weights}
    w_{i,j} = 
    \begin{bmatrix}
        \omega_{1}u_{1}^{\gamma}(x_i,t_j) & \dots & \omega_{n}u_{n}^{\gamma}(x_i,t_j)
    \end{bmatrix}\t,
\end{eqnarray}
for $\gamma\geq0$ and $\omega_1,\dots,\omega_n\in\mathbb{R}$ where $n$ is the dimensionality of the system. Note that (i) noiseless data are modeled by letting $\omega_1,\dots,\omega_n=0$, (ii) constant-variance error used in ordinary least squares is modeled by letting $\gamma=0$, $\omega_1,\dots,\omega_n=1$, and (iii) non-constant-variance error (e.g. proportional error) used in generalized least squares is modeled by letting $\gamma>0$, $\omega_1,\dots,\omega_n\neq0$. Therefore, to account for the statistical error model in Eq~\eqref{eq:statistical_model}, the GLS objective function 
\begin{eqnarray}
\label{eq:gls_loss}
        \mathcal{L}_{\text{GLS}} = \frac{1}{MN} \sum_{i=1,j=1}^{M,N} \left[ \frac{u_{\text{MLP}}(x_i, t_j) - u_{i,j}}{\left| u_{\text{MLP}}(x_i, t_j) \right|^{\gamma}} \right]^2,
\end{eqnarray}
is used with proportionality constant $\gamma=0.2$. Note that $\gamma$ was tuned numerically following the methodology suggested in~\cite{lagergren_2020} (see Methods Section for more details). 

%\noindent\textbf{PDE Loss.}
The next term $\mathcal{L}_{\text{PDE}}$ ensures $u_{\text{MLP}}$ satisfies the solution of the governing PDE. For ease of notation, let $\hat{u}_{i,j} \equiv u_{\text{MLP}}(x_i,t_j)$, $\hat{D}_{i,j} \equiv D_{\text{MLP}}(u_{\text{MLP}}(x_i,t_j))$, and $\hat{G}_{i,j} \equiv G_{\text{MLP}}(u_{\text{MLP}}(x_i,t_j))$. Then for the reaction-diffusion equation, the error term takes the following form:
\begin{eqnarray}
\label{eq:pde_loss}
    \mathcal{L}_{\text{PDE}} = \frac{1}{MN} \sum_{i=1,j=1}^{M,N} \Bigg[ \underbrace{\vphantom{\Bigg[}\frac{\partial \hat{u}_{i,j}}{\partial t}}_{\text{LHS}} - \underbrace{\vphantom{\Bigg[}\left(\frac{\partial}{\partial x} \left( \hat{D}_{i,j} \, \frac{\partial \hat{u}_{i,j}}{\partial x} \right) + \hat{G}_{i,j} \, \hat{u}_{i,j}\right)}_{\text{RHS}}  \Bigg]^2,
\end{eqnarray}
where LHS and RHS denote the left-hand- and right-hand-sides of the governing PDE, respectively. Thus, by driving $\mathcal{L}_{\text{PDE}}$ to zero, the RHS is trained to match the LHS. Through this process, the nonlinear forms of $D_{\text{MLP}}$ and $G_{\text{MLP}}$ are learned despite not being directly observed. See the Methods Section for additional implementation details, including a random sampling procedure that enforces this PDE constraint everywhere in the input domain during training.

%\noindent\textbf{Biological Constraint Loss.} 
Biological information and domain expertise are incorporated into the BINNs framework by adding penalties in the loss term $\mathcal{L}_{\text{Constr}}$. For the reaction-diffusion equation, the diffusivity and growth rates are assumed to be within biologically feasible ranges $[D_{\text{min}}, D_{\text{max}}]$ and $[G_{\text{min}}, G_{\text{max}}]$, respectively. Further, diffusion is also assumed to be non-decreasing and growth to be non-increasing with respect to cell density. The corresponding constraints take the form:
\begin{align}
    \label{eq:constr_loss}
    \mathcal{L}_{\text{Constr}} = \frac{1}{MN} \Bigg[ \sum_{\substack{i=1,j=1 \\ \hat{D}<D_{\text{min}} \\ \hat{D}>D_{\text{max}}}}^{M,N} \left(\hat{D}_{i,j}\right)^2 & + \sum_{\substack{i=1,j=1 \\ \partial \hat{D}/\partial \hat{u}<0}}^{M,N} \left(\frac{\partial \hat{D}_{i,j}}{\partial \hat{u}_{i,j}}\right)^2 
    \\
    & + \sum_{\substack{i=1,j=1 \\ \hat{G}<G_{\text{min}} \\ \hat{G}>G_{\text{max}}}}^{M,N} \left(\hat{G}_{i,j}\right)^2 + \sum_{\substack{i=1,j=1 \\ \partial \hat{G}/\partial \hat{u}>0}}^{M,N} \left(\frac{\partial \hat{G}_{i,j}}{\partial \hat{u}_{i,j}}\right)^2 \Bigg]. \nonumber
\end{align}
The maximum and minimum diffusivity and growth rates considered in~\cite{simpson_2016} were used to force $D_{\text{MLP}}$ and $G_{\text{MLP}}$ to stay within biologically realistic ranges. The constraints on $D_{\text{MLP}}$ and $G_{\text{MLP}}$ shown in Eq~\eqref{eq:constr_loss} were used for all computational results in this work. See the Methods Section for biological motivations and numerical implementation details of these constraints. 

%\subsection*{Numerically solving PDEs with Biologically-informed neural networks}\label{sec:pdesolve_binns}
\subsection*{Evaluation procedure}

Because the model prediction $u_{MLP}(x,t)$ is only a surrogate model for the dynamical system $u(x,t)$, it is possible that this approximation may contain errors, particularly in areas where the PDE constraint given by Eq~\eqref{eq:pde_loss} is not satisfied. To ensure that the inferred diffusion and growth terms lead to biologically realistic dynamics, the reaction-diffusion equation given by Eq~\eqref{eq:reaction_diffusion} is solved numerically with a method-of-lines approach using $\mathcal{D} = D_{\text{MLP}}$ and $\mathcal{G} = G_{\text{MLP}}$. Note that this model is well-defined because $D_{\text{MLP}}$ and $G_{\text{MLP}}$ are continuously differentiable functions of the cell density, $u$. \JHL{Further, BINNs are retrained multiple times for each data set in which the forward simulation using the learned PDE terms that yields the smallest GLS error (Eq~\eqref{eq:gls_loss}) is saved.} All fits to the data shown in the Results Section are numerical solutions to the PDE in Eq~\eqref{eq:reaction_diffusion} using the learned diffusivity and growth functions. See the Methods Section for numerical implementation details of the PDE forward solver. 

\section*{Results}

%The Results section should provide details of all of the experiments that are required to support the conclusions of the paper. There is no specific word limit for this section, but details of experiments that are peripheral to the main thrust of the article and that detract from the focus of the article should not be included. The section may be divided into subsections, each with a concise subheading. The section should be written in the past tense.

\subsection*{Simulation case study}

Since the diffusivity and growth terms are inferred by BINNs through learning $D_{\text{MLP}}$ and $G_{\text{MLP}}$, respectively, the ability of BINNs to learn biologically accurate representations of these terms must first be tested. To investigate this, data were simulated using the classical FKPP and Generalized Porous-FKPP equations with parameter values from~\cite{simpson_2016} for the scratch assay data with initial cell density 20,000 cells per well. Additionally, the simulated data were obscured with artificial observation error using the statistical model in Eq~\eqref{eq:statistical_model} with $\gamma=0.2$. Each simulation used the initial condition from the scratch assay data with initial cell density 20,000 cells per well. Using the same level of sparsity (i.e. 37 spatial points and five time points), the BINNs framework was shown to (i) approximate the dynamical system accurately and (ii) approximate the general forms of the diffusivity and growth terms. See~\nameref{sim_model_fit} and~\nameref{sim_parameter_fit} for the model and parameter fits, respectively. \JHL{This case study demonstrates that BINNs are able to learn accurate representations of the diffusivity and growth functions from biologically realistic noisy sparse data, however, further analysis, like model selection and comparison, is omitted here and instead explored using experimental data.}

\subsection*{Reaction-diffusion BINNs for experimental data}

\begin{figure}[!h]
    \centering
    \includegraphics[width=1.0\textwidth]{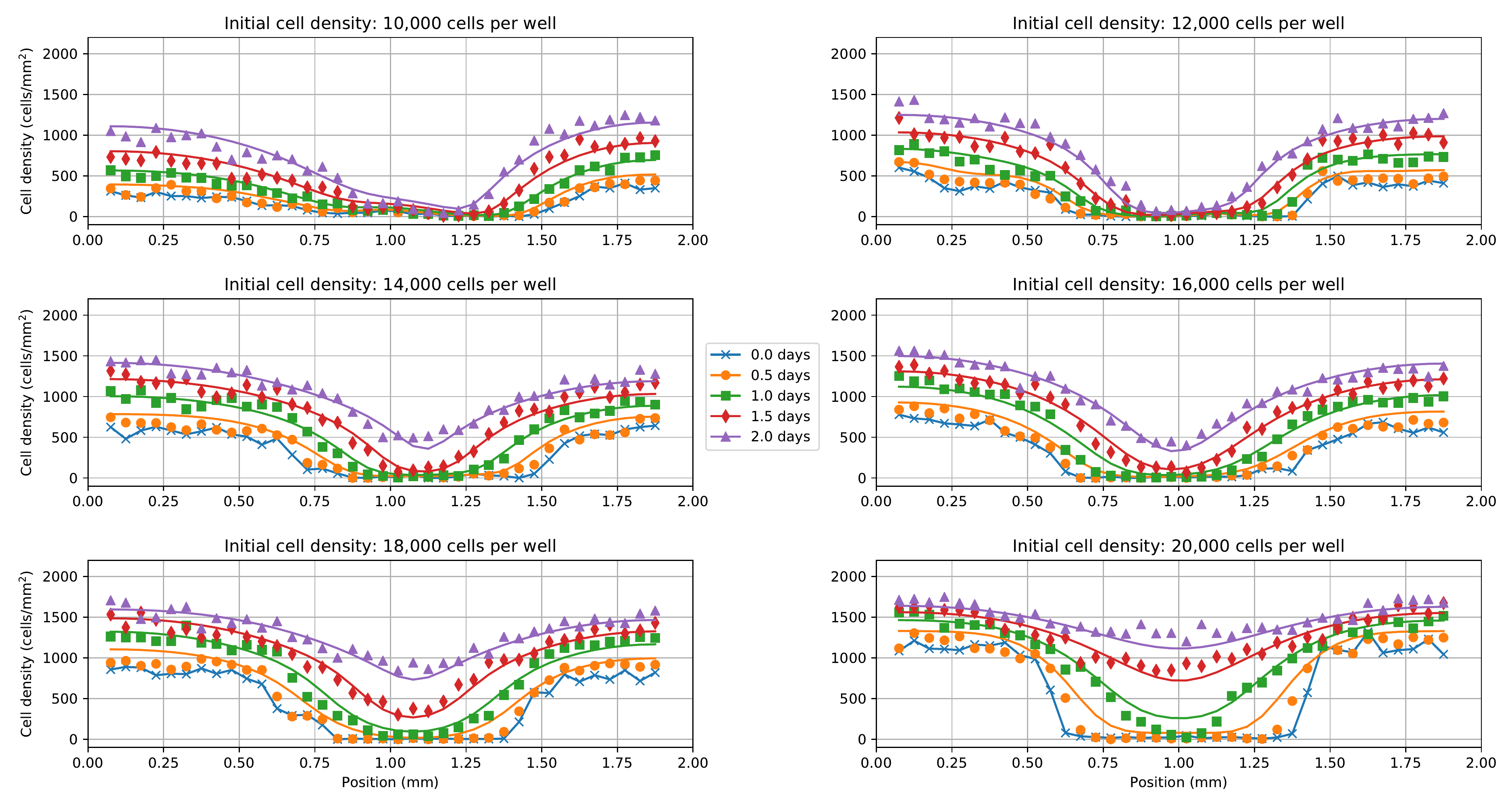}
    \caption{\textbf{Reaction-diffusion BINN solutions.} Predicted cell density profiles using BINNs with the governing reaction-diffusion PDE in Eq~\eqref{eq:reaction_diffusion_binn}. Each subplot corresponds to an experiment with a different initial cell density (i.e. 10,000, 12,000, 14,000, 16,000, 18,000, and 20,000 cells per well). Solid lines represent the numerical solution to Eq~\eqref{eq:reaction_diffusion_binn} using $D_{\text{MLP}}$ and $G_{\text{MLP}}$. The markers represent the experimental scratch assay data. }
    \label{fig:rd_binn_results}
\end{figure}

As described in the previous sections, the diffusivity and growth functions are approximated by deep neural networks, $\mathcal{D} = D_{\text{MLP}}(u)$ and $\mathcal{G} = G_{\text{MLP}}(u)$, resulting in a governing PDE of the form
\begin{eqnarray}
\label{eq:reaction_diffusion_binn}
    u_t = (D_{\text{MLP}}(u)u_x)_x + G_{\text{MLP}}(u)u,
\end{eqnarray}
where $D_{\text{MLP}}$ and $G_{\text{MLP}}$ are functions of the cell density $u$. A BINN was trained for each data set with varying initial cell density. The resulting numerical PDE solutions using the trained $D_{\text{MLP}}$ and $G_{\text{MLP}}$ are shown in Fig~\ref{fig:rd_binn_results}.

\begin{figure}[!h]
    \centering
    \includegraphics[width=1.0\textwidth]{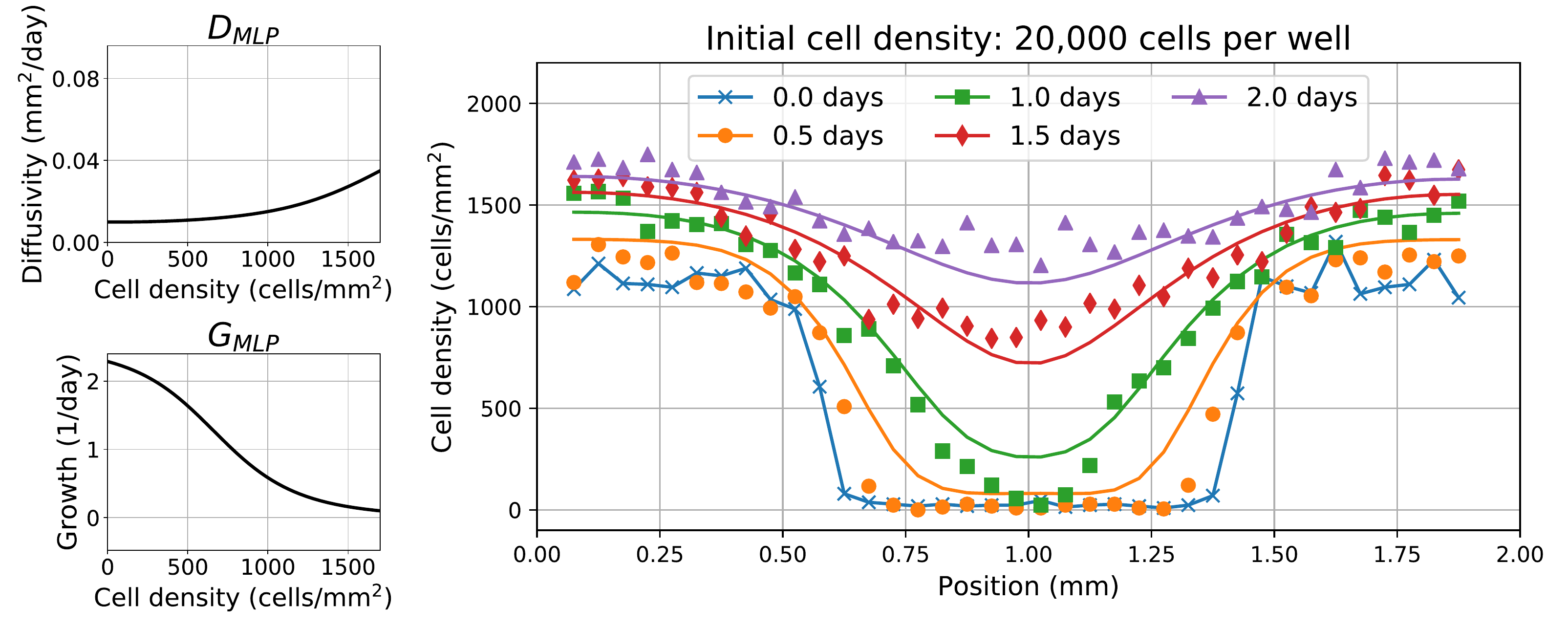}
    \caption{\textbf{Reaction-diffusion BINN terms and discrepancy.} Left: learned diffusivity and growth functions, $D_{\text{MLP}}$ and $G_{\text{MLP}}$, evaluated over cell density, $u$. Right: Predicted cell density profiles using BINNs with the governing reaction-diffusion PDE in Eq~\eqref{eq:reaction_diffusion_binn} for data with initial cell density 20,000 cells per well. Solid lines represent the numerical solution to Eq~\eqref{eq:reaction_diffusion_binn} using $D_{\text{MLP}}$ and $G_{\text{MLP}}$. The markers represent the experimental scratch assay data.}
    \label{fig:rd_binn_discrepancy}
\end{figure}

While the model fits shown in Fig~\ref{fig:rd_binn_results} are excellent for lower initial cell densities, there still remains a significant amount of model discrepancy at higher initial cell densities. \JHL{GLS residual errors were computed to provide an additional way of visualizing the model discrepancy (see~\nameref{rdbinn_residuals}) in which non-i.i.d. residuals are clearly present at higher initial cell densities. To investigate the specific form of the model discrepancy,} Fig~\ref{fig:rd_binn_discrepancy} shows the learned diffusivity and growth functions with the corresponding model fit for the data set with an initial cell density of 20,000 cells per well. 

Fig~\ref{fig:rd_binn_discrepancy} reveals clear model discrepancy in two main areas: (i) at high cell densities (i.e. $x\in[0,0.25]$ mm and $x\in[1.75,2.0]$ mm for $t\in[0,1]$ days) where diffusion is negligible and the dynamics are governed primarily by growth; and (ii) at low cell densities (i.e. $x\in[0.5,1.5]$ mm for $t\in[0,1]$ days) where growth is negligible and the dynamics are primarily governed by diffusion. In particular, the discrepancy is largest for early time points where the diffusion and growth dynamics appear too rapid. \JHL{The solutions of $D_{\text{MLP}}$ and $G_{\text{MLP}}$ are also qualitatively similar to the classical FKPP equation in which the learned diffusivity function is relatively constant while the learned growth function is approximately linearly decreasing with cell density, $u$. However, despite $D_{\text{MLP}}$ and $G_{\text{MLP}}$ learning biologically realistic functions for the diffusivity and growth, the persistent model discrepancy observed across multiple data sets with high initial cell densities (see Fig~\ref{fig:rd_binn_results}) suggests that the reaction-diffusion equation described in Eq~\eqref{eq:reaction_diffusion_binn} may be insufficient to fully capture the underlying dynamics of cell migration for these data.} From a mathematical modeling perspective, the model discrepancy at early time points suggests the existence of a time delay that scales the magnitude of the density-dependent diffusion and growth rates. Biological reasons behind this phenomenon may include cell damage from the scratch assay protocol or changes in cell functions where more cells become immobile/non-proliferative as the cell density approaches carrying capacity~\cite{dydowiczova_2020, poumay_1995, neurohr_2020}. See the Discussion Section for more details. 

\subsection*{Delay-reaction-diffusion BINNs for experimental data}

\begin{figure}[!h]
    \centering
    \includegraphics[width=1.0\textwidth]{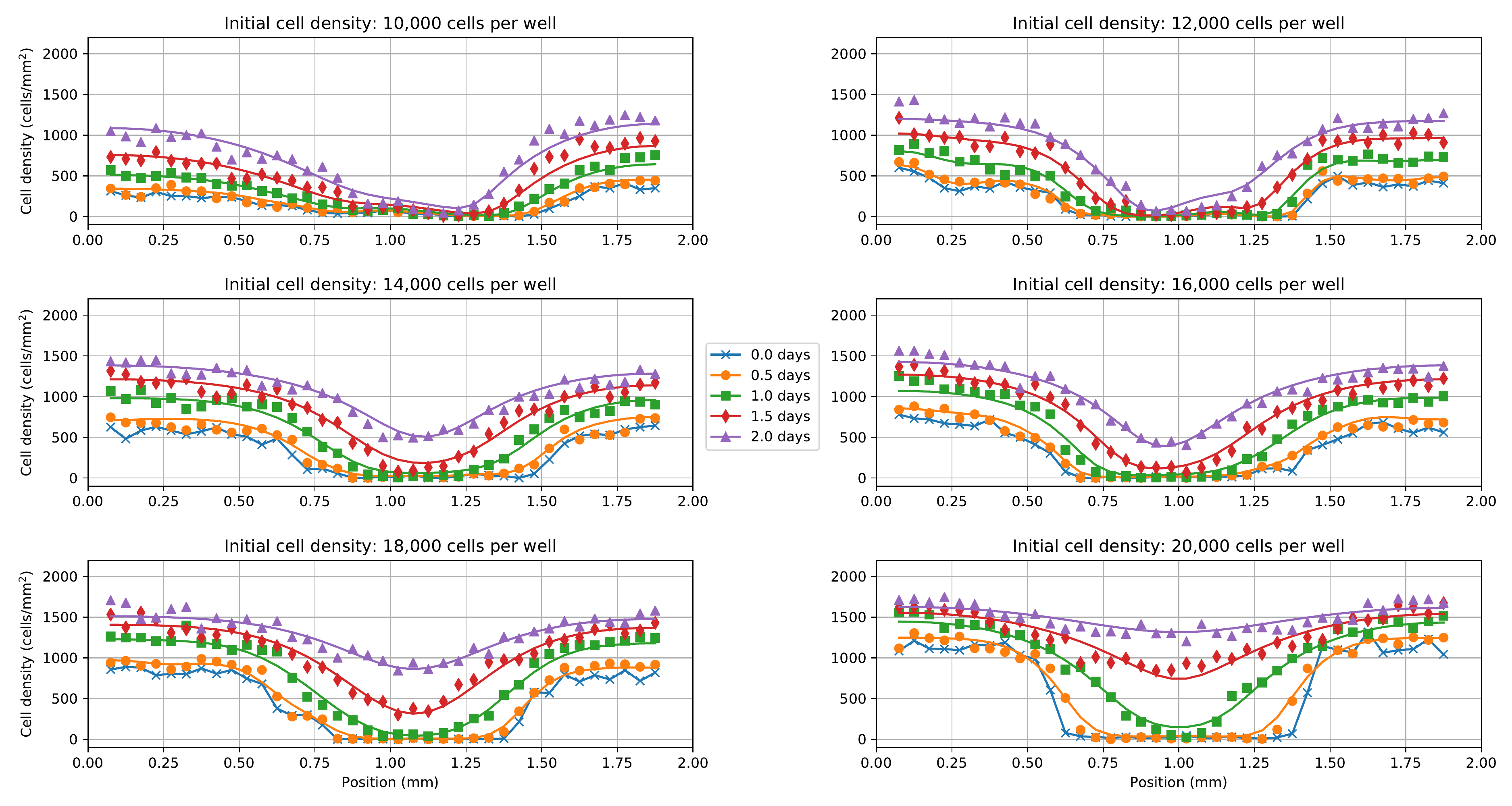}
    \caption{\textbf{Delay-reaction-diffusion BINN solutions.} Predicted cell density profiles using BINNs with the governing delay-reaction-diffusion PDE in Eq~\eqref{eq:delay_reaction_diffusion_binn}. Each subplot corresponds to an experiment with a different initial cell density (i.e. 10,000, 12,000, 14,000, 16,000, 18,000, and 20,000 cells per well). Solid lines represent the numerical solution to Eq~\eqref{eq:delay_reaction_diffusion_binn} using $T_{\text{MLP}}$, $D_{\text{MLP}}$, and $G_{\text{MLP}}$. The markers represent the experimental scratch assay data.}
    \label{fig:delay_rd_binn_results}
\end{figure}

Motivated by the model discrepancy for data sets with high initial cell density, the reaction-diffusion equation in Eq~\eqref{eq:reaction_diffusion_binn} was modified by including a time delay described by an additional neural network function $T_{\text{MLP}}(t)$. The new term $T_{\text{MLP}}(t)$ is a continuously differentiable function of time that is constrained to be non-decreasing and output values between 0 and 1. In this way $T_{\text{MLP}}$ can scale the strength of the density-dependent diffusivity and growth terms in time. Letting the diffusivity, $\mathcal{D}$, and growth, $\mathcal{G}$, terms of the governing PDE be functions of $u$ and $t$, they are replaced with $\mathcal{D} = T_{\text{MLP}}(t)D_{\text{MLP}}(u)$ and $\mathcal{G} = T_{\text{MLP}}(t)G_{\text{MLP}}(u)$. This results in a governing PDE of the form
\begin{eqnarray*}
    u_t = \big(T_{\text{MLP}}(t)D_{\text{MLP}}(u)u_x\big)_x + T_{\text{MLP}}(t)G_{\text{MLP}}(u)u,
\end{eqnarray*}
which simplifies to 
\begin{eqnarray}
\label{eq:delay_reaction_diffusion_binn}
    u_t = T_{\text{MLP}}(t)\Big((D_{\text{MLP}}(u)u_x)_x + G_{\text{MLP}}(u)u\Big),
\end{eqnarray}
where $D_{\text{MLP}}$ and $G_{\text{MLP}}$ are functions of the cell density $u$ and $T_{\text{MLP}}$ is a function of time $t$. Note that $T_{\text{MLP}}$ was chosen to be separable from $D_{\text{MLP}}$ and $G_{\text{MLP}}$ since the density-dependent dynamics of diffusion and growth are assumed to be consistent throughout time. \JHL{Further, it was assumed that both $D_{\text{MLP}}$ and $G_{\text{MLP}}$ are scaled by the same time delay;} see the Discussion Section for more details. BINNs governed by the PDE in Eq~\eqref{eq:delay_reaction_diffusion_binn} were trained for each data set with varying initial cell density. The resulting forward simulations using the trained $T_{\text{MLP}}$, $D_{\text{MLP}}$, and $G_{\text{MLP}}$ networks are shown in Fig~\ref{fig:delay_rd_binn_results}.

\begin{figure}[!h]
    \centering
    \includegraphics[width=1.0\textwidth]{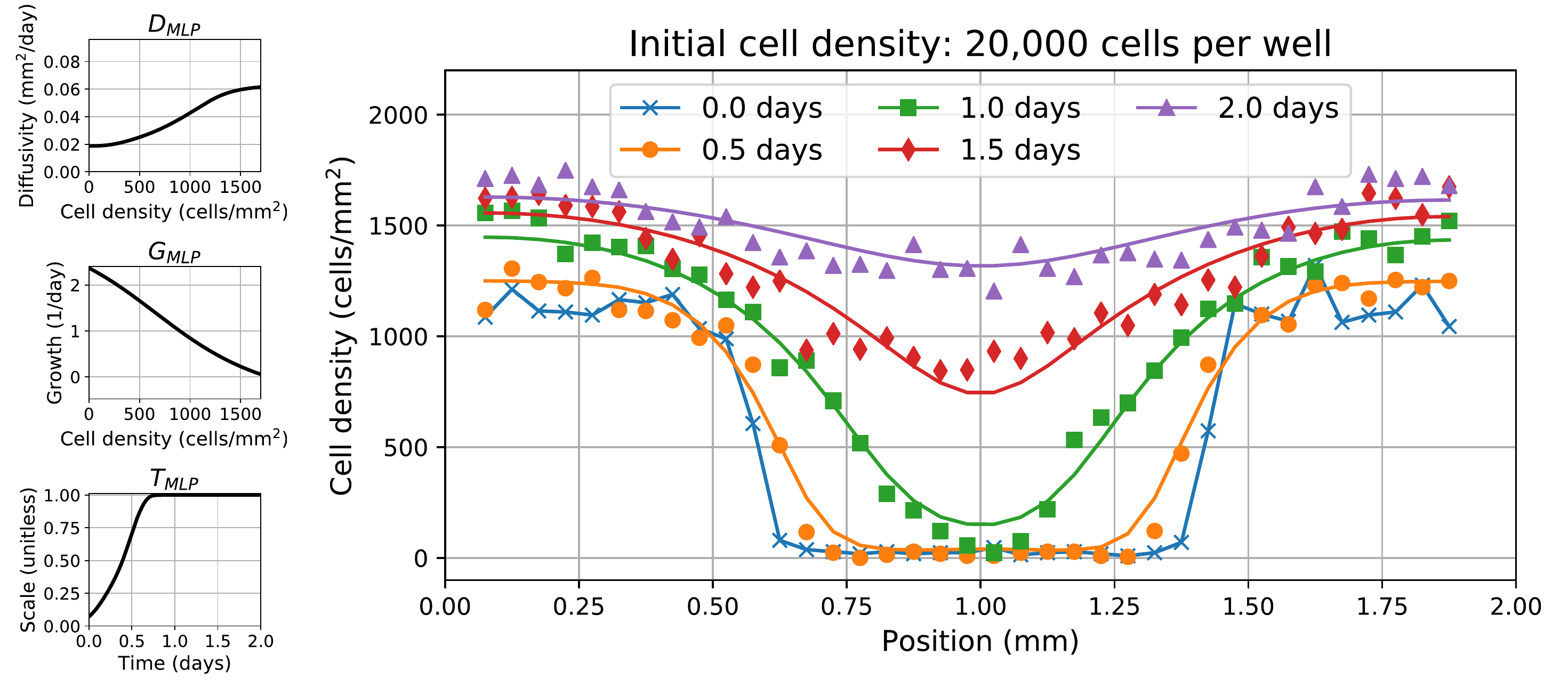}
    \caption{\textbf{Delay-reaction-diffusion BINN terms and discrepancy.} Left: learned diffusivity and growth functions, $D_{\text{MLP}}$ and $G_{\text{MLP}}$, evaluated over cell density, $u$, and delay function, $T_{\text{MLP}}$, evaluated over time, $t$. Right: Predicted cell density profiles using BINNs with the governing delay-reaction-diffusion PDE in Eq~\eqref{eq:delay_reaction_diffusion_binn} for data with initial cell density 20,000 cells per well. Solid lines represent the numerical solution to Eq~\eqref{eq:delay_reaction_diffusion_binn} using $D_{\text{MLP}}$, $G_{\text{MLP}}$, and $T_{\text{MLP}}$. The markers represent the experimental scratch assay data.}
    \label{fig:delay_rd_binn_discrepancy}
\end{figure}

The model fits shown in Fig~\ref{fig:delay_rd_binn_results} demonstrate that virtually all of the model discrepancy across each initial cell density was removed by including a time delay. \JHL{This is confirmed further using GLS residual errors (see~\nameref{drdbinn_residuals}) where the residuals are approximately i.i.d. even at higher initial cell densities. Similar to the reaction-diffusion case,} Fig~\ref{fig:delay_rd_binn_discrepancy} shows the learned diffusivity, growth, and delay functions with the corresponding model fit for the data set with initial cell density of 20,000 cells per well. 

Fig~\ref{fig:delay_rd_binn_discrepancy} shows that the model discrepancy in areas with high and low cell densities at early time points has been practically eliminated. This is most clearly seen in the delay-reaction-diffusion model solution at the second time point (i.e. $t=0.5$ days), which matches the data more accurately than the reaction-diffusion model in Eq~\eqref{eq:reaction_diffusion_binn} at the same time point (see Fig~\ref{fig:rd_binn_discrepancy}). Moreover, $D_{\text{MLP}}$ and $G_{\text{MLP}}$ for the delay-reaction-diffusion BINN learned similar forms of the diffusivity and growth compared to the reaction-diffusion case. However, the delay term $T_{\text{MLP}}$ reveals that the diffusion and growth dynamics described by $D_{\text{MLP}}$ and $G_{\text{MLP}}$ are scaled down for early time points (i.e. $t<1$) before $T_{\text{MLP}}$ converges to 1, allowing $D_{\text{MLP}}$ and $G_{\text{MLP}}$ to come into full effect. \JHL{This observation is of particular importance since the majority of scratch assay data are reported within this time delay region (i.e. 4, 6, 12, or 24 hrs)~\cite{aoki_propagating_2017, arciero_continuum_2011, bindschadler_sheet_2007, matsubayashi_erk_2004}. Importantly, not accounting for a time delay within this region may potentially explain why scratch assay experiments are notoriously difficult to reproduce~\cite{simpson_2016}.}

\subsection*{Guided mechanistic model selection}

\begin{figure}[!h]
    \centering
    \includegraphics[width=1.0\textwidth]{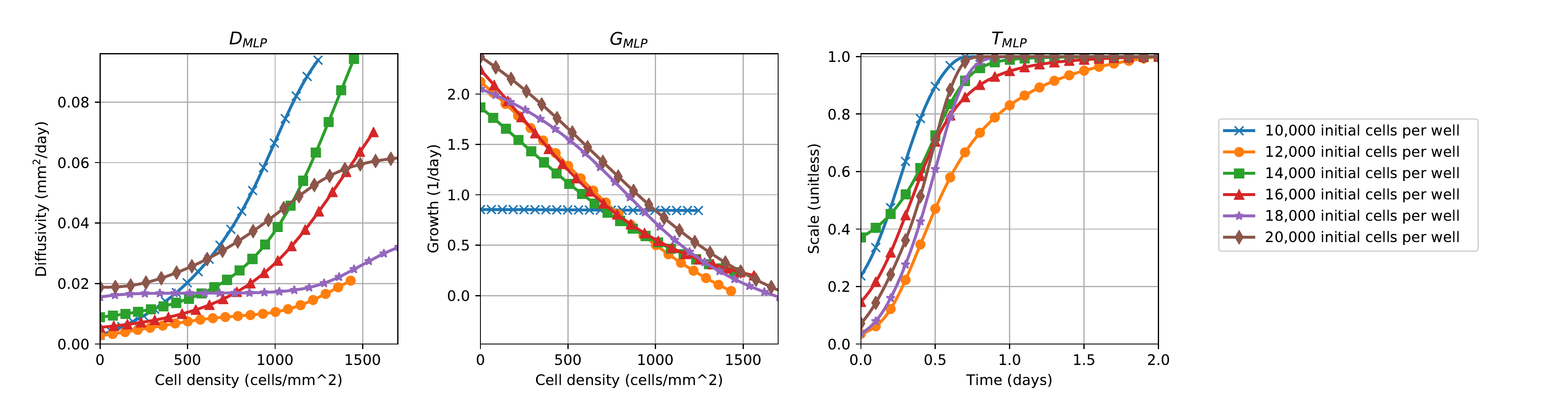}
    \caption{\textbf{Delay-reaction-diffusion BINN terms.} The learned diffusivity, $D_{\text{MLP}}$, growth, $G_{\text{MLP}}$, and delay, $T_{\text{MLP}}$, functions extracted from the corresponding BINNs with governing delay-reaction-diffusion PDE in Eq~\eqref{eq:delay_reaction_diffusion_binn}. Each line corresponds to an experiment with a different initial cell density (i.e. 10,000, 12,000, 14,000, 16,000, 18,000, and 20,000 cells per well). Note that $D_{\text{MLP}}$ and $G_{\text{MLP}}$ have different lengths since they are evaluated between the minimum and maximum observed cell densities corresponding to each data set.}
    \label{fig:delay_rd_binn_parameters}
\end{figure}

The diffusion, growth, and delay networks $D_{\text{MLP}}$, $G_{\text{MLP}}$, and $T_{\text{MLP}}$ were used to guide the selection of biologically realistic mechanistic models for downstream use in a traditional mathematical modeling framework. Each network solution corresponding to the six scratch assay data sets is shown in Fig~\ref{fig:delay_rd_binn_parameters}. 

 From Fig~\ref{fig:delay_rd_binn_parameters}, the learned diffusivities for each experiment with different initial cell density are non-zero when $u=0$ and appear increasing and concave up (like power laws) with respect to the cell density $u$. On the other hand, the learned growth terms are approximately linear, which is consistent with logistic models, and the learned delay terms all exhibit sigmoidal dynamics. Note that the outlying $G_{\text{MLP}}$ solution for the scratch assay data set with 10,000 initial cells per well is likely an artifact of the observed cell densities in that experiment not approaching the carrying capacity, and therefore leading to unrealistic learned dynamics. \JHL{Based on qualitative analysis of these plots}, the following mechanistic delay-reaction-diffusion equation is proposed to satisfy each scratch assay data set:
\begin{eqnarray}
\label{eq:mechanistic_delay_reaction_diffusion}
    u_t &= \mathcal{T}(t)\Big((\mathcal{D}(u)u_x)_x + \mathcal{G}(u)u\Big),
\end{eqnarray}
with diffusivity, growth, and delay functions
\begin{subequations}
    \begin{align}
        \mathcal{D} &= D_0 + D\left(\frac{u}{K}\right)^m, \label{eq:mechanistic_diffusion} \\
        \mathcal{G} &= ru\left(1-\frac{u}{K}\right), \label{eq:mechanistic_growth} \\
        \mathcal{T} & = \frac{1}{1+e^{-(\beta_1t + \beta_0)}}, \label{eq:mechanistic_delay}
    \end{align}
\end{subequations}
respectively. The diffusivity function $\mathcal{D}$ in Eq~\eqref{eq:mechanistic_diffusion} is a combination of the classical FKPP and Generalized Porous-FKPP diffusivity function, with baseline cell diffusivity $D_0$, diffusion coefficient $D$, and exponent $m$. The growth function $\mathcal{G}$ in Eq~\eqref{eq:mechanistic_growth} is chosen to be the logistic growth function with intrinsic growth rate $r$ and carrying capacity $K$. The delay function $\mathcal{T}$ in Eq~\eqref{eq:mechanistic_delay} is represented by the logistic regression function with parameters $\beta_0$ and $\beta_1$. One advantage of using a mathematical model with specified functional forms and parameters described by Eq~\eqref{eq:mechanistic_diffusion}-\eqref{eq:mechanistic_delay} is that standard parameter estimation techniques can now be used. This enables a comparison of the BINN-guided model in Eq~\eqref{eq:mechanistic_delay_reaction_diffusion} to other mechanistic models, namely, the classical FKPP and Generalized Porous-FKPP equations. 

\subsection*{Model comparison}

\begin{figure}[!h]
    \centering
    \includegraphics[width=1.0\textwidth]{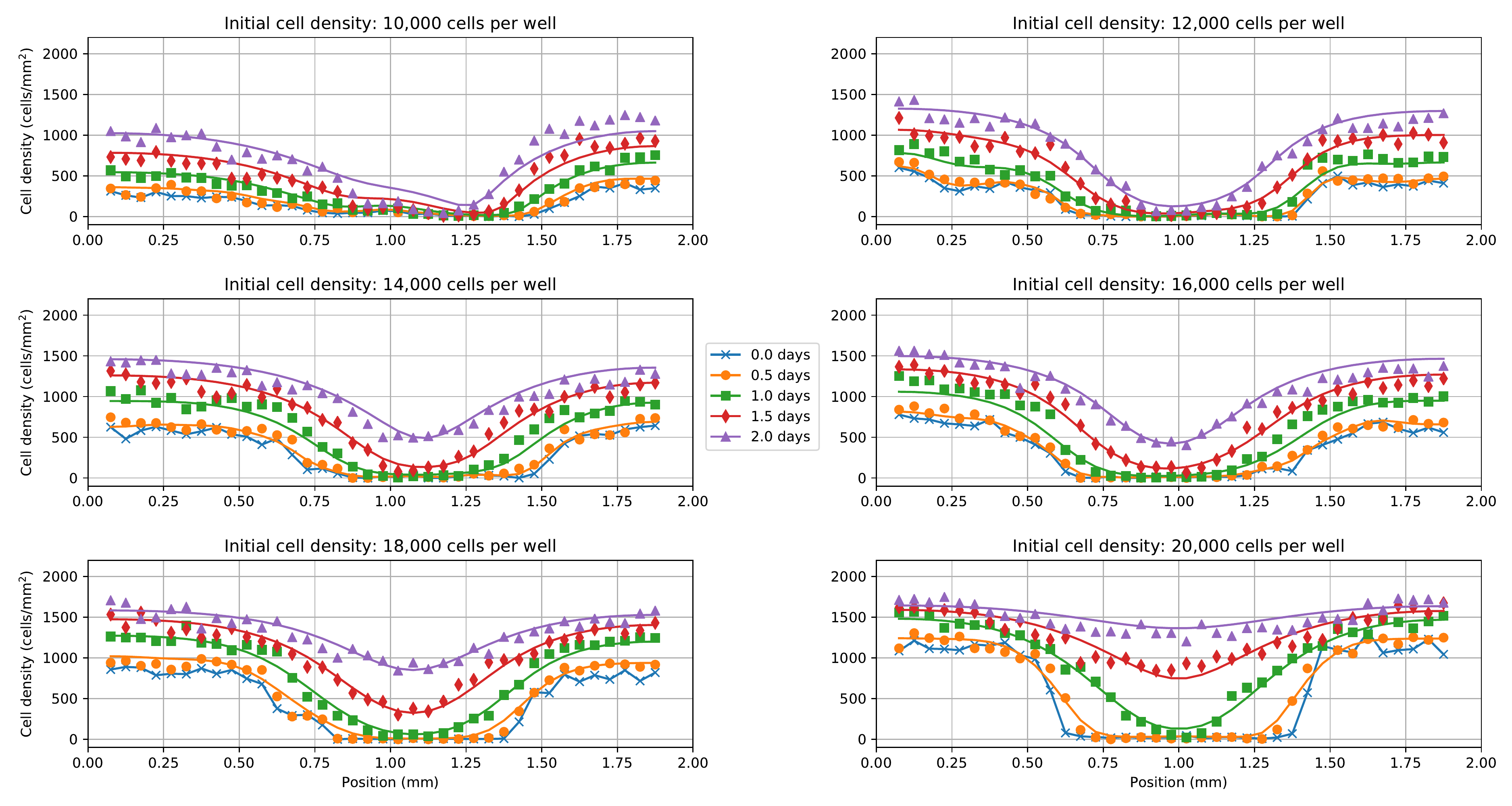}
    \caption{\textbf{BINN-guided delay-reaction-diffusion model solutions.} Predicted cell density profiles using the delay-reaction-diffusion model in Eq~\eqref{eq:mechanistic_delay_reaction_diffusion}. Each subplot corresponds to an experiment with a different initial cell density (i.e. 10,000, 12,000, 14,000, 16,000, 18,000, and 20,000 cells per well). Solid lines represent the numerical solution to Eq~\eqref{eq:mechanistic_delay_reaction_diffusion} using the parameters that minimize $\mathcal{L}_{\text{GLS}}$ in Eq~\eqref{eq:gls_loss}. The markers represent the experimental scratch assay data.}
    \label{fig:mechanistic_fkpp}
\end{figure}

The BINN-guided delay-reaction-diffusion model in Eq~\eqref{eq:mechanistic_delay_reaction_diffusion} was compared to the classical FKPP equation
\begin{eqnarray}
\label{eq:classical_fkpp}
    u_t &= (Du_{x})_{x} + ru\left(1-\frac{u}{K}\right),
\end{eqnarray}
with diffusion coefficient $D$, intrinsic growth rate $r$, and carrying capacity $K$ and Generalized Porous-FKPP equation
\begin{eqnarray}
\label{eq:porous_fkpp}
    u_t &= (D(\frac{u}{K})^m u_x)_x + ru\left(1-\frac{u}{K}\right),
\end{eqnarray}
with additional exponent $m$. These models were used as a baseline for comparison since they have been identified as the current state-of-the-art in modeling these data~\cite{simpson_2016, warne_using_2019}. The parameters of each model were optimized numerically using the generalized least squares error function in Eq~\eqref{eq:gls_loss} with the adjusted statistical error model in Eq~\eqref{eq:statistical_model} with $\gamma=0.2$. Note that the carrying capacity was fixed at $K=1.7\times10^3$ $\nicefrac{\text{cells}}{\text{mm}^2}$ and not optimized because it was empirically validated in~\cite{simpson_2016}. The resulting model fits and parameter values for the classical FKPP and Generalized Porous-FKPP models are shown in~\nameref{fkpp_solutions},~\nameref{pfkpp_solutions},~\nameref{fkpp_parameters}, and~\nameref{pfkpp_parameters}. The solutions of the BINN-guided delay-reaction-diffusion model in Eq~\eqref{eq:mechanistic_delay_reaction_diffusion} to each data set are shown in Fig~\ref{fig:mechanistic_fkpp}.

The predicted cell density profiles in Fig~\ref{fig:mechanistic_fkpp} closely matched the scratch assay data which suggests that the proposed model in Eq~\eqref{eq:mechanistic_delay_reaction_diffusion} with Eqs~\eqref{eq:mechanistic_diffusion}-\eqref{eq:mechanistic_delay} successfully captured the learned dynamics from $T_{\text{MLP}}$, $D_{\text{MLP}}$, and $G_{\text{MLP}}$. The optimized parameter values across each data set are shown in Table~\ref{tab:dpfkpp_parameters}. Note that the parameters were rescaled to $\mu\text{m}$ and hours (hr) for comparison with~\cite{simpson_2016} and~\cite{warne_using_2019}. 

\begin{table} [!h]
    \centering
    \caption{\textbf{BINN-guided delay-reaction-diffusion model parameters.}}
    \begin{tabular}{l|cccccc}
        \multicolumn{1}{l}{} & \multicolumn{6}{c}{Initial cell density} \\
        Parameter & 10,000 & 12,000 & 14,000 & 16,000 & 18,000 & 20,000
        \\ \hline
        $D_0$ ($\nicefrac{\mu\text{m}^2}{\text{hr}}$) & 95.7  & 353.3 & 482.1 & 604.3 & 804.0 & 675.8 \\
        $D$ ($\nicefrac{\mu\text{m}^2}{\text{hr}}$) & 3987.1 & 3166.4 & 3775.0 & 3773.8 & 2201.8 & 1954.9 \\
        $m$ (unitless) & 1.5976 & 3.4708 & 1.9060 & 3.5173 & 3.2204 & 0.9876 \\
        $r$ ($\nicefrac{1}{\text{hr}}$) & 0.0525 & 0.0714 & 0.0742 & 0.0798 & 0.0772 & 0.0951 \\
        $\beta_0$ (unitless) & -1.0292 & -3.3013 & -3.1953 & -2.9660 & -1.2695 & -4.0651 \\
        $\beta_1$ ($\nicefrac{1}{\text{hr}}$) & 0.2110 & 0.2293 & 0.2761 & 0.2180 & 0.1509 & 0.4166 \\
    \end{tabular}
    \begin{flushleft} 
        Table of model parameters for Eq~\eqref{eq:mechanistic_delay_reaction_diffusion} calibrated for each scratch assay data set. Each column corresponds to an experiment with different initial cell density (i.e. 10,000, 12,000, 14,000, 16,000, 18,000, and 20,000 cells per well). 
    \end{flushleft}
    \label{tab:dpfkpp_parameters}
\end{table}

%$D_0$ (mm$^2$/day)   & 0.0023 & 0.0085 & 0.0116 & 0.0145 & 0.0193 & 0.0162 \\
%$D$ (mm$^2$/day)     & 0.0957 & 0.0760 & 0.0906 & 0.0906 & 0.0528 & 0.0469 \\
%$m$ (unitless)       & 1.5976 & 3.4708 & 1.9060 & 3.5173 & 3.2204 & 0.9876 \\
%$r$ (1/day)          & 1.2611 & 1.7135 & 1.7801 & 1.9144 & 1.8524 & 2.2825 \\
%$\beta_0$ (unitless) & -1.0292 & -3.3013 & -3.1953 & -2.9660 & -1.2695 & -4.0651 \\
%$\beta_1$ (1/day)    & 5.0638 & 5.5041 & 6.6262 & 5.2324 & 3.6211 & 9.9986 \\

\JHL{Table~\ref{tab:dpfkpp_parameters} reveals that many of the parameters relating to density-dependent diffusion and growth show trends (e.g. $D_0$ and $r$ increasing) with initial cell density similar to~\cite{simpson_2016}. The implications of this observation are considered in the Discussion Section.} To compare the three models quantitatively, the generalized least squares (GLS) errors were computed for each model and data set and reported in Table~\ref{tab:gls_errors}.

\begin{table} [!h]
    \centering
    \caption{\textbf{Generalized least squares (GLS) errors.}}
    \begin{tabular}{l|cccccc}
        \multicolumn{1}{l}{} & \multicolumn{6}{c}{Initial cell density} \\
        Model & 10,000 & 12,000 & 14,000 & 16,000 & 18,000 & 20,000
        \\ \hline
        classical FKPP & 786.80 & 557.28 & 616.76 & 619.12 & 685.17 & 964.19 \\ 
        Porous-FKPP & 681.18 & 540.29 & 418.57 & 566.89 & 744.44 & 928.38 \\ 
        BINN-guided model & \textbf{557.01} & \textbf{317.18} & \textbf{410.79} & \textbf{393.15} & \textbf{307.74} & \textbf{386.52} 
    \end{tabular}
    \begin{flushleft} 
        Table of GLS errors between the model solutions and scratch assay data. Each column corresponds to an experiment with different initial cell density (i.e. 10,000, 12,000, 14,000, 16,000, 18,000, and 20,000 cells per well). Bold numbers represent the minimum GLS error across the three models.
    \end{flushleft}
    \label{tab:gls_errors}
\end{table}

The results in Table~\ref{tab:gls_errors} showed that Eq~\eqref{eq:mechanistic_delay_reaction_diffusion} with Eqs~\eqref{eq:mechanistic_diffusion}-\eqref{eq:mechanistic_delay} fit each data set more accurately than the classical FKPP or Generalized Porous-FKPP models. This behavior is not surprising given that the BINN-guided model is more complex. Therefore, model selection methods, which balance model accuracy with model complexity, were also used to compare the \emph{quality} of each model relative to the others. In particular, the modified Akaike Information Criterion (AIC) from~\cite{banks_2017} was used to account for the statistical error model in Eq~\eqref{eq:statistical_model}. See Table~\ref{tab:aic_scores} for the AIC scores across each model and data set.

\begin{table} [!h]
    \centering
    \caption{\textbf{Akaike Information Criterion (AIC) scores.}}
    \begin{tabular}{l|cccccc}
        \multicolumn{1}{l}{} & \multicolumn{6}{c}{Initial cell density} \\
        Model & 10,000 & 12,000 & 14,000 & 16,000 & 18,000 & 20,000
        \\ \hline
        classical FKPP & 1239.6 & 1175.8 & 1194.5 & 1195.2 & 1214.0 & 1277.2 \\
        Porous-FKPP & 1214.9 & 1172.0 & \textbf{1124.8} & 1180.9 & 1231.3 & 1272.2 \\
        BINN-guided model & \textbf{1183.7} & \textbf{1079.5} & 1127.3 & \textbf{1119.2} & \textbf{1073.9} & \textbf{1116.1} 
    \end{tabular}
    \begin{flushleft} 
        Table of AIC scores for each model and scratch assay data set. Each column corresponds to an experiment with different initial cell density (i.e. 10,000, 12,000, 14,000, 16,000, 18,000, and 20,000 cells per well). Bold numbers represent the minimum AIC score across the three models.
    \end{flushleft}
    \label{tab:aic_scores}
\end{table}

The results in Table~\ref{tab:aic_scores} showed that the BINN-guided delay-reaction-diffusion model outperforms the classical FKPP and Generalized Porous-FKPP models across all data sets except with initial cell density 14,000 cells per well. This discrepancy follows from Table~\eqref{eq:gls_loss} where the additional parameters in Eq~\eqref{eq:mechanistic_delay_reaction_diffusion} only slightly decreased the GLS error for the data set with initial density of 14,000. Finally, to quantify the ``value'' of adding the novel delay term in Eq~\eqref{eq:mechanistic_delay} the differences between AIC scores for each model and the minimum AIC score, denoted by $\Delta$AIC, are shown in Table~\ref{tab:daic_scores}.

\begin{table}[!h]
    \centering
    \caption{\textbf{Difference Akaike Information Criterion ($\Delta$AIC) scores.}}
    \begin{tabular}{l|cccccc}
        \multicolumn{1}{l}{} & \multicolumn{6}{c}{Initial cell density} \\
        Model & 10,000 & 12,000 & 14,000 & 16,000 & 18,000 & 20,000
        \\ \hline
        classical FKPP & 55.90 & 96.26 & 69.71 & 76.01 & 140.08 & 161.11 \\
        Porous-FKPP & 31.23 & 92.54 &  0.00 & 61.70 & 157.42 & 156.11 \\
        BINN-guided model &  0.00 &  0.00 &  2.53 &  0.00 &  0.00 &  0.00 
    \end{tabular}
    \begin{flushleft} 
        Table of AIC differences ($\Delta$AIC) between each model and scratch assay data set. Each column corresponds to an experiment with different initial cell density (i.e. 10,000, 12,000, 14,000, 16,000, 18,000, and 20,000 cells per well). Each $\Delta$AIC score represents the difference between a model's AIC score and the minimum recorded AIC score for that data set.
    \end{flushleft}
    \label{tab:daic_scores}
\end{table}

Table~\ref{tab:daic_scores} suggests that the delay term is most impactful for data sets with large initial density (i.e. 18,000 and 20,000 cells per well) since the $\Delta$AIC scores are significantly larger for these data sets. Biological analysis and explanations for these results are considered in the following Discussion Section.

\section*{Discussion}

%The Discussion should spell out the major conclusions of the work along with some explanation or speculation on the significance of these conclusions. How do the conclusions affect the existing assumptions and models in the field? How can future research build on these observations? What are the key experiments that must be done?

In this work, biologically-informed neural networks (BINNs) were introduced as a flexible and robust equation learning method for real-world biological applications. The BINNs framework was demonstrated using experimental biological data from scratch assays~\cite{simpson_2016} and used to discover a delay term that had not yet been considered in the modeling of these data. The trained diffusivity, growth, and delay networks were used to guide the selection of the mechanistic model in Eq~\eqref{eq:mechanistic_delay_reaction_diffusion} with Eqs~\eqref{eq:mechanistic_diffusion}-\eqref{eq:mechanistic_delay}, which was shown to model the data more accurately than the current state-of-the-art models (i.e. classical FKPP and Generalized Porous-FKPP equations). The results shown in this work suggest that the BINNs framework can be successfully applied to a wide range of biological and physical problems where the data are sparse and the governing dynamics are unknown. The biological motivations for various aspects of the BINNs framework and significance of the results are discussed in the following paragraphs.

% biological motivations for time delay
The model solutions in Fig~\ref{fig:rd_binn_results} and Fig~\ref{fig:rd_binn_discrepancy} indicated that using only density-dependent diffusivity and growth functions $\mathcal{D}(u)$ and $\mathcal{G}(u)$ was not sufficient to fully capture the scratch assay dynamics. Fig~\ref{fig:rd_binn_discrepancy} highlighted this discrepancy at the second time measurement ($t=0.5$ days) in which the model failed to capture the areas of both high and low cell density, despite letting $\mathcal{D}$ and $\mathcal{G}$ be universal function-approximating neural networks. In particular, the model solutions in the areas of high cell density (i.e. $x\in[0.0, 0.5]$ and $x\in[1.5, 2.0]$) showed exponential convergence to the carrying capacity, which successfully captured the data for later time points ($t\geq1$ days) but over-predicted for early time points ($t<1$ days). Similarly, diffusion in areas of low cell density (i.e. $x\in[0.5, 1.5]$) over-predicted the cell density profile for early time points but then matched the data accurately for later time points. From a mathematical perspective, this motivates the existence of a time delay that scales the density-dependent dynamics to be reduced for early time points and larger for later time points. There are also several biological motivations for considering a time delay. For example,~\cite{dydowiczova_2020} showed how cells are damaged at the borders of the scratch as a result of the experimental scratch assay protocol. Cell damage can potentially inhibit the communication between cells and physically block healthy cells from diffusing into uncolonized spatial regions. Another source of delay may stem from changes in density-dependent cell functions (e.g. differentiation, division, and senescence). Studies have shown that cells are more likely to terminally differentiate when cell populations approach carrying capacity~\cite{poumay_1995, neurohr_2020}. Therefore, scratch assay experiments that are performed for high density populations may contain fewer mobile/proliferative cells at the borders of the scratch, thus causing a time delay in the cell migration dynamics.

% time delay function details
A general framework for incorporating the delay term may be to consider diffusivity and growth functions $\mathcal{D} = D_{\text{MLP}}(u,t)$ and $\mathcal{G} = G_{\text{MLP}}(u,t)$, respectively. However, since the dynamics of diffusion and growth are assumed to be consistent throughout time, the diffusion and growth terms were chosen to be separable functions composed of diffusivity $\mathcal{D}(u)$, growth $\mathcal{G}(u)$, and delay $\mathcal{T}(t)$. Additionally, it was assumed that both diffusion and growth were scaled by the same time delay $\mathcal{T}(t)$ as opposed to a diffusion delay $\mathcal{T}_{\text{D}}(t)$ and growth delay $\mathcal{T}_{\text{G}}(t)$. This assumption may not be accurate if the time delay is a result of density-dependent changes in cell function where cells become mobile and proliferative at different rates. \JHL{In particular, since migration and proliferation have very different timescales, it might be natural to expect that the delays would also have different timescales.} However, since the numerical solutions using $\mathcal{T}(t)$ matched the data sufficiently accurately, this question is left for future work. Finally, $\mathcal{T}(t)$ was constrained to output values between 0 and 1 and forced to be increasing with time. These constraints were chosen to ensure that the delay term modeled the time-dependent changes in cell dynamics for early time points but converged to unity by later time points. 

% binn guided mechanistic model discussion
In this work, BINNs revealed that the reaction-diffusion system in Eq~\eqref{eq:reaction_diffusion} with cell density-dependent diffusivity and growth functions was insufficient to capture the data dynamics. However, the model discrepancy for data sets with large initial cell density motivated the development of a time delay which significantly improved the model accuracy and resolved the observed discrepancy. The diffusivity, growth, and delay networks were used to posit a mechanistic model (i.e. Eq~\eqref{eq:mechanistic_delay_reaction_diffusion} with Eqs~\eqref{eq:mechanistic_diffusion}-\eqref{eq:mechanistic_delay}). Using the logistic growth model (Eq~\eqref{eq:mechanistic_growth}) for the growth function and logistic regression (Eq~\eqref{eq:mechanistic_delay}) for the delay function followed straightforwardly from the parameter network solutions in Fig~\ref{fig:delay_rd_binn_parameters}, however, the diffusivity function in Eq~\eqref{eq:mechanistic_diffusion} warrants further discussion. 

\JHL{Opinions vary between the biological validity of (i) the classical FKPP and (ii) the Generalized Porous-FKPP diffusivity functions. For example, one study compared (i) and (ii) using experimental wound size data and found that (ii) with $m=4$ provided the best fit to the data~\cite{sherratt_1990}. Another study fit (i) and (ii) to experimental cell migration data with different cell populations and found that one population was best described by constant diffusivity in (i) and the other by nonlinear diffusivity with $m=1$ in (ii)~\cite{sengers_2007}. These studies do not reveal which approach is best, but they demonstrate that care is warranted.} The diffusivity network ($D_{\text{MLP}}$) solutions showed significant variability across the scratch assay data sets, so the posited mechanistic model was chosen to respect the observed variability while also being as simple as possible. Therefore, Eq~\eqref{eq:mechanistic_diffusion} was chosen to be a combination of the diffusivity functions in (i) and (ii). This way both (i) and (ii) can be seen as nested models of the posited diffusivity function by setting either $D=0$ or $D_0=0$. Yet the posited diffusivity is still simple, as it only increases the number of parameters with respect to (ii) by one. It may be the case that the true diffusivity function is even more complex, such as a linear combination of powers:
\[
    \mathcal{D} = D_0 + D_1\left(\frac{u}{K}\right)^{m_1} + D_2\left(\frac{u}{K}\right)^{m_2},
\]
with baseline diffusivity $D_0$, diffusion rates $D_1$ and $D_2$, carrying capacity $K$, and exponents $m_1$ and $m_2$. However, these considerations are beyond the scope of the present work and left for future work.

% analysis of optimized parameters
The parameters of (i) the classical FKPP in Eq~\eqref{eq:classical_fkpp}, (ii) the Generalized Porous-FKPP in Eq~\eqref{eq:porous_fkpp}, and (iii) the BINN-guided model in Eq~\eqref{eq:mechanistic_delay_reaction_diffusion} with Eqs~\eqref{eq:mechanistic_diffusion}-\eqref{eq:mechanistic_delay} were optimized numerically for each scratch assay data set. The optimized parameters for (i) in~\nameref{fkpp_parameters} all fall within the ranges reported in~\cite{simpson_2016}. However, this is not the case for any set of parameter values for (ii) as shown in~\nameref{pfkpp_parameters}. This is likely due to the parameter optimization being conducted using the adjusted statistical error model in Eq~\eqref{eq:statistical_model} with $\gamma=0.2$ and since the exponent $m$ in the Porous-FKPP diffusivity function was not fixed at $m=1$ as in~\cite{simpson_2016}. However, in both (i) and (ii), the diffusion coefficient, $D$, and intrinsic growth rate, $r$, showed variability with initial cell density, similar to the conclusions drawn in~\cite{simpson_2016}. Therefore, in theory, if the delay term in Eq~\eqref{eq:mechanistic_delay} accounts for the time it takes for density-dependent growth and diffusion to become active in the system, which may be a function of initial cell density, then the variability among diffusion coefficients and intrinsic growth rates for the BINN-guided delay-reaction-diffusion model should be reduced across the scratch assay experiments. However, from the optimized parameter values in Table~\ref{tab:dpfkpp_parameters}, the baseline diffusion rate $D_0$ and intrinsic growth rate $r$ generally increase with initial cell density and the diffusion coefficient $D$ generally decreases with initial cell density. This observation may indicate (i) practical identifiability issues between the diffusion, growth, and delay terms or \JHL{(ii) the existence additional mechanisms that are not accounted for in the model. To confirm this, a Bayesian parameter estimation framework can be used to examine practical identifiability of parameters \cite{lagergren_forecasting_2018, adoteye_correlation_2015}.} Then, a possible strategy to mitigate this issue would be to optimize the parameters of Eq~\eqref{eq:mechanistic_delay_reaction_diffusion} with Eqs~\eqref{eq:mechanistic_diffusion} and \eqref{eq:mechanistic_growth} jointly across each scratch assay data set while allowing the delay parameters in Eq~\eqref{eq:mechanistic_delay} to be tuned separately for each set. This exploration is left for future work.

% model accuracy and AIC scores
The BINN-guided delay-reaction-diffusion model was compared to the baseline classical FKPP and Generalized Porous-FKPP models using both GLS errors and modified AIC scores. The GLS errors in Table~\ref{tab:gls_errors} showed that the BINN-guided model fits the data more accurately than the baseline models across each scratch assay data set. However, this improvement in accuracy is due to the increased model complexity (i.e. number of parameters and PDE terms) in the BINN-guided model. Therefore, to rank the quality of each model, AIC scores were also computed since they balance model accuracy with model complexity. The AIC scores reported in Table~\ref{tab:aic_scores} indicate that the BINN-guided model also exceeds the baseline models in terms of relative quality across each scratch assay data set except with initial cell density 14,000 cells per well, in which the Generalized Porous-FKPP model has a slightly smaller AIC score. In other words, Tables~\ref{tab:gls_errors} and~\ref{tab:aic_scores} indicate that the BINN-guided model performs as well or better than the state-of-the-art in modeling the suite of scratch assay experiments from~\cite{simpson_2016}. In particular, this advantage is afforded by including the delay term in Eq~\eqref{eq:mechanistic_delay}. To quantify the relative value of adding the delay term, the AIC scores from Table~\ref{tab:aic_scores} are used to compute difference AIC ($\Delta$AIC) scores in Table~\ref{tab:daic_scores} in which the $\Delta$AIC score for a fixed model and data set is given by the difference between the corresponding AIC score and the minimum AIC score across all models for the given data set. The $\Delta$AIC scores in Table~\ref{tab:daic_scores} indicate that the relative value of the delay term is largest for data sets with initial cell density 18,000 and 20,000 cells per well. This observation is supported by the relevant biology discussed at the beginning of this section, in which large initial cell densities either (i) result in more damaged cells near the borders of the scratch, (ii) cause more cells in the population to have terminally differentiated away from mobile/proliferative cell functions, or (iii) some combination of (i) and (ii) and other potentially unconsidered biological sources, all of  which increase the potential time delay before the density-dependent diffusion and growth dynamics become the primary drivers of the temporal evolution of the system. 

\subsection*{Conclusions and future work}

BINNs, a robust and flexible framework for equation learning with sparse and noisy data, was demonstrated and used to posit a mechanistic equation that outperforms the state-of-the-art in modeling experimental scratch assay data. The development, training, and evaluation of BINNs and the resulting model selection and analysis were reported to justify these claims. \JHL{The discovered time delay term may have important implications for the reproducibility and modeling of scratch assays, since the majority of the reported data fall within the time delay region.} Some of the drawbacks of the BINNs method and opportunities for future work and development are discussed below.

Since BINNs rely on multilayer perceptrons (MLPs), the learned dynamics may not generalize well outside the training domain. For example, in the present work, if the observed cell densities for a particular experiment do not approach the carrying capacity (e.g. the scratch assay data set with 10,000 initial cells per well) then the learned dynamics given by $D_{\text{MLP}}$ and $G_{\text{MLP}}$ may lead to biologically unrealistic behavior (see $G_{\text{MLP}}$ solutions in Fig~\ref{fig:delay_rd_binn_parameters}). Further, since none of the scratch assay data reported values that significantly exceeded the empirically set carrying capacity, $G_{\text{MLP}}$ would likely not generalize well to a scenario with exceedingly large observed cell densities. Options for mitigating this issue include (i) replacing unrealistic MLP terms with mechanistic models (e.g. logistic growth instead of $G_{\text{MLP}}$) if the particular dynamics are known \textit{a priori}, or (ii) adding additional constraints which force the MLP terms to satisfy specific values (e.g. $G_{\text{MLP}}(u=K)=0$). 

An opportunity for future development is quantifying the uncertainty of both the approximate solution $u_{\text{MLP}}$ and the parameter networks $D_{\text{MLP}}$, $G_{\text{MLP}}$, and $T_{\text{MLP}}$. From the frequentist perspective, so called ``subagging'' (i.e. subsample aggregating) can be used to build posterior distributions of the model solutions and parameter networks~\cite{buhlmann_2012}. In this framework, one simply samples $N$ training/validation splits and trains a BINN for each split. Then kernel density estimation or some other equivalent methodology can be used to build distributions from the $N$ number of trained BINNs. Alternatively, from the Bayesian perspective, physics-informed neural networks were recently extended to Bayesian physics-informed neural networks (B-PINNs)~\cite{yang_2020}. In this framework, Bayesian neural networks are substituted for $u_{\text{MLP}}$ and regularized using a pre-specified governing PDE. In the BINNs framework, Bayesian neural networks could also be substituted for $D_{\text{MLP}}$, $G_{\text{MLP}}$, and $T_{\text{MLP}}$ to quantify the uncertainty of the PDE terms in addition to the model solution.

While BINNs were demonstrated using one-dimensional reaction-diffusion PDEs for scratch assay data in this work, they can be applied on a wide spectrum of physical and biological problems (for both ODE and PDE systems) in which the governing dynamics are unknown and highly nonlinear. A straightforward next step for this work would be to evaluate BINNs on the two-dimensional scratch assay image data that were used to construct the one-dimensional cell density profiles in~\cite{simpson_2016}. Further, more complicated cell dynamics could be incorporated into the governing system in the present work by including PDE terms that describe cell population heterogeneity or additional biological mechanisms for damaged (but not dead) cells at the borders of the scratch. 

\JHL{BINNs were used to address a canonical problem in the field of collective cell migration by analyzing how the combination of density-dependent cell motility and proliferation drive the temporal dynamics of cell invasion during an experimental scratch assay. This novel framework revealed new mechanistic and biological insights into this process by guiding the derivation of a mathematical model that has not been considered previously using traditional mathematical modeling approaches. The classical FKPP and Generalized Porous-FKPP models are ubiquitous in modeling cell migration and proliferation, yet the BINNs methodology presented here revealed that these models may fail to incorporate all of the relevant mechanisms underlying this process. These results suggest that new models incorporating a time delay may be necessary to accurately capture the dynamics within the first day of a scratch assay, i.e., just after the scratch is introduced. Based on the success of this work, BINNs establish a new paradigm for data-driven equation learning from sparse and noisy data that could enable the rapid development and validation of mathematical models for a broad range of real-world applications throughout biology including ecology, epidemiology, and cell biology.}

\section*{Methods}

%The Materials and Methods should provide enough detail for reproduction of the findings. Submit detailed protocols for newer or less established methods. Well-established protocols may simply be referenced.

All methods herein were implemented in Python 3.6.8 using the PyTorch 1.2.0 deep learning library. All data and code are made publicly available at \url{https://github.com/jlager}. The following section is intended to make BINNs feasible for a wide range of biological applications. In particular, this section covers (i) the importance of data pre-processing, (ii) strategies for using real-world knowledge to design effective neural network models, (iii) the complete training protocol ranging from selecting appropriate statistical error models and hyperparameters to balancing the multi-objective error function, and (iv) numerical implementation details for forward solving BINNs-guided PDEs. 

\subsection*{Data pre-processing}

Input and output standardization are common practice to stabilize neural network training~\cite{theodoridis_pattern_2009}. Since the scratch assay data in~\cite{simpson_2016} reported cell densities on the order of $u=\mathcal{O}(10^{-3})$ $\nicefrac{\text{cells}}{\mu\text{m}^2}$ at spatial locations on the order of $x=\mathcal{O}(10^{3})$ $\mu\text{m}$ for time points on the order of $t=\mathcal{O}(10)$ hours, these variables needed to be standardized. Without standardization, the neural network models failed to converge for these data because (i) the network inputs ($x$ and $t$) differed by several orders of magnitude from each other and (ii) the network inputs ($x$ and $t$) and outputs ($u$) also differed by several orders of magnitude. By rescaling $x$ and $t$ to millimeters (mm) and days, respectively, the adjusted variables ranged from $x=\mathcal{O}(1)$ mm, $t=\mathcal{O}(1)$ days, and cell density $u=\mathcal{O}(10^3)$ $\nicefrac{\text{cells}}{\text{mm}^2}$. Standardizing $x$ and $t$ addressed (i) while (ii) is addressed by using scaling factors discussed in the following section. The cell density profile at the left boundary was removed since it was consistently larger than the remaining cell densities across all six data sets.

\subsection*{Network design}

BINNs are centered around $u_{\text{MLP}}$, a function-approximating multilayer perceptron (MLP) (also known as an artificial neural network). MLPs, like polynomials~\cite{stone_generalized_1948}, are in the class of \emph{universal function approximators}, meaning that they can approximate any continuous bounded functions on a closed interval arbitrarily well under some reasonable assumptions~\cite{hornik_1991}. For the scratch assay data in the present work, $u_{\text{MLP}}$ inputs spatiotemporal vectors $\boldsymbol{x} = \begin{bmatrix}x, t\end{bmatrix}$ and outputs the corresponding approximations to the cell density $u$. To give $u_{\text{MLP}}$ sufficient capacity to approximate the solution to the governing PDE, the network is chosen to have three hidden layers with 128 neurons in each layer, resulting in a model with approximately 30,000 total parameters. Concretely, $u_{\text{MLP}}$ takes the form
\begin{eqnarray}
\label{eq:u_mlp}
    u_{\text{MLP}}(\boldsymbol{x}) = \alpha \cdot \phi\Big(\,\sigma\Big(\,\sigma\big(\,\sigma(\,\boldsymbol{x}W_1 + b_1\,)\,W_2 + b_2\,\big)\,W_3 + b_3\,\Big)\,W_4 + b_4\,\Big),
\end{eqnarray}
where the trainable parameters $W_i$ and $b_i$ denote weight matrices and bias vectors for the $i^{\text{th}}$ layer, $\sigma(\cdot)$ and $\phi(\cdot)$ denote nonlinear activation functions, and $\alpha$ denotes a scaling factor. Each hidden layer uses a ``sigmoid'' activation function (i.e. $\sigma(x) = 1/(1+e^{-x})$) while the output layer uses a ``softplus'' activation function (i.e. $\phi(x) = \ln(1+e^x)$). The softplus activation function is a particular design choice since it is a continuously differentiable function that forces the predicted cell densities to be non-negative, and has been previously shown to be well-suited for biological transport models~\cite{lagergren_2020}. Finally, to account for the difference in scale between the inputs ($x,t=\mathcal{O}(1)$) and outputs ($u=\mathcal{O}(10^3)$), the MLP outputs are post-multiplied by the experimentally validated carrying capacity (i.e. $\alpha=1.7\times10^{3}$) from~\cite{simpson_2016}. Note that in practice, if values like this are unknown, one can simply let $\alpha$ be the maximum observed cell density or some other similar quantity. The key here is to ensure the orders of magnitude between the network inputs and outputs are similar so that the parameters of the MLP do not have to account for the change of scale~\cite{theodoridis_pattern_2009}.

The diffusivity, growth, and delay functions of the governing PDEs are modeled with neural networks $D_{\text{MLP}}(u_{\text{MLP}})$ and $G_{\text{MLP}}(u_{\text{MLP}})$, and $T_{\text{MLP}}(t)$. All three MLPs share the same number of layers as $u_{\text{MLP}}$ but use 32 neurons per layer. These networks are chosen to be smaller for both computational efficiency and because the parameter dynamics are assumed to be simpler than the cell density dynamics $u$. The hidden layers use sigmoid activation functions. The output layer for $D_{\text{MLP}}$ uses a softplus activation because diffusion is assumed to be non-negative for all cell densities. Since the growth term can be negative (e.g. logistic growth when the cell density exceeds the carrying capacity), a linear output (i.e. no activation function) is used in the final layer for $G_{\text{MLP}}$. The output layer for $T_{\text{MLP}}$ uses the sigmoid function to constrain the outputs to $(0, 1)$. Finally, as with $u_{\text{MLP}}$, the inputs and outputs of $D_{\text{MLP}}$ and $G_{\text{MLP}}$ are also standardized. In particular, the inputs of both networks (i.e. $u_{\text{MLP}}$) are divided by the carrying capacity $K=1.7\times10^{3}$ while the outputs of $D_{\text{MLP}}$ are multiplied by 0.096 $\nicefrac{\text{mm}^2}{\text{day}}$ and the outputs of $G_{\text{MLP}}$ are multiplied by 2.4 $\nicefrac{1}{\text{day}}$. These values were the maximum diffusion and growth values considered in~\cite{simpson_2016}. Similar to $u_{\text{MLP}}$, the input and output scaling factors ensure the MLP parameters do not have to account for changes in scale. No standardization was used for $T_{\text{MLP}}$ since its inputs and outputs are of the same order (i.e. $\mathcal{O}(1)$).

\subsection*{Training procedure}

The BINN parameters (i.e. weights and biases of $u_{\text{MLP}}$, $D_{\text{MLP}}$, $G_{\text{MLP}}$, and $T_{\text{MLP}}$) are optimized using the first-order gradient-based Adam optimizer~\cite{kingma_adam_2017} with default hyper-parameters and minibatch-optimization. To prevent over-fitting, the scratch assay data were randomly partitioned into 80\%/20\% training and validation sets. The network parameters were updated iteratively to minimize $\mathcal{L}_{\text{Total}}$ in Eq~\eqref{eq:total_error} on the training set and saved on \emph{relative} improvement in validation error. In other words, the model parameters were saved if the relative difference between (i) the validation error in the current iteration and (ii) the smallest recorded validation error exceeded $5\%$. Finally, since the parameters of each BINN are randomly initialized and applied to different data sets, early stopping of 5,000 (i.e. training was stopped if the relative validation error did improve for 5,000 consecutive epochs) was used to guarantee the convergence of each BINN independently. The implementation details of each term in $\mathcal{L}_{\text{Total}}$ (i.e. $\mathcal{L}_{\text{GLS}}$, $\mathcal{L}_{\text{PDE}}$, and $\mathcal{L}_{\text{Constr}}$) are discussed in more detail below.

The first term $\mathcal{L}_{\text{GLS}}$ in Eq~\eqref{eq:gls_loss} corresponds to the generalized least squares (GLS) distance between $u_{\text{MLP}}$ and the observation data $u_{i,j}$. Since the error process is assumed to be i.i.d., the parameters of the statistical model in Eq~\eqref{eq:statistical_model} (i.e. $\gamma$) must first be calibrated. Following~\cite{lagergren_2020}, $u_{\text{MLP}}$ is trained using $\mathcal{L}_{\text{GLS}}$ as an objective function for $\gamma=0.0, 0.2, 0.4, 0.6$ (recall that $\gamma=0.0$ represents the ordinary least squares case) for each data set. After qualitative assessment of the modified residual errors (see~\nameref{stat_model_selection}), $\gamma=0.2$ was identified as the value that produced the most i.i.d. residuals across each of the six data sets. Using the calibrated statistical error model, $\mathcal{L}_{\text{GLS}}$ is evaluated at each training iteration using mini-batches (i.e. randomly selected subsets) of input/output data. In general, using a small batch size acts as an additional form of regularization that helps neural networks escape local minima during training and allows for better generalization~\cite{keskar_large-batch_2017}. However, this significantly increases the computational cost of training due to the increased number of training iterations needed to converge. Therefore, BINNs were trained using mini-batches of size 37 (i.e. $1/4$ the number of points in the training set) which was found to balance the accuracy and computational cost. 

To ensure $u_{\text{MLP}}$ satisfies the solution of the governing PDE, the terms $\mathcal{L}_{\text{PDE}}$ in Eq~\eqref{eq:pde_loss} and $\mathcal{L}_{\text{Constr}}$ in Eq~\eqref{eq:constr_loss} are included in $\mathcal{L}_{\text{Total}}$ as a form of regularization. However, since the scratch assay data are sparse, simply training $u_{\text{MLP}}$ using $\mathcal{L}_{\text{Total}}$ at the observed data locations can result in unrealistic dynamics in between data points. Therefore, to ensure $u_{\text{MLP}}$ satisfies the solution of a governing PDE \emph{everywhere} in the input domain, $\mathcal{L}_{\text{PDE}}$ and $\mathcal{L}_{\text{Constr}}$ are evaluated at 10,000 uniformly randomly sampled points $x_i\in[x_{\text{min}}, x_{\text{max}}]$ and $t_j\in[t_{\text{min}}, t_{\text{max}}]$ at \emph{each} training iteration. Without the random sampling procedure, $u_{\text{MLP}}$ can severely overfit to the data. To illustrate the importance of the random sampling procedure, the model fits, GLS errors, and PDE errors are shown in~\nameref{pde_random_sampling} for three cases in which (i) no PDE regularization is used, (ii) PDE regularization is used at the data locations, and (iii) PDE regularization is used at 10,000 randomly sampled points. In particular,~\nameref{pde_random_sampling} shows that in option (i) $u_{\text{MLP}}$ overfits the data practically everywhere in the input domain, (ii) $u_{\text{MLP}}$ overfits everywhere except at the data locations (see vertical lines in third subplot of row b), and (iii) the random sampling procedure results in the smallest amount of PDE error and the largest amount of GLS error. The desired behavior is shown in option (iii) since $u_{\text{MLP}}$ fits the data as accurately as allowed by the governing PDE.

The third error term $\mathcal{L}_{\text{Constr}}$ constrains $D_{\text{MLP}}$, $G_{\text{MLP}}$, and $T_{\text{MLP}}$ to exhibit biologically realistic values and dynamics. Choosing appropriate constraints can be ambiguous when the relevant literature gives conflicting suggestions. For example, when designing a derivative constraint for the diffusivity network $D_{\text{MLP}}$,~\cite{anguige_one-dimensional_2009} suggest that diffusion should decrease with cell density due to cell-to-cell adhesion whereas~\cite{nardini_modeling_2016} suggest the opposite in which cells promote the migration of others. To mitigate this, BINNs were trained without any constraints on $D_{\text{MLP}}$ and $G_{\text{MLP}}$ in order to visualize the collective behavior of the parameter networks (see~\nameref{unconstr_binn_parameters}). Note that $T_{\text{MLP}}$ was still forced to be non-decreasing. The network evaluations in~\nameref{unconstr_binn_parameters} showed unrealistic parameter dynamics for some data sets, but their collective behavior was used to design derivative constraints that forced $D_{\text{MLP}}$ to increase as a function of cell density and $G_{\text{MLP}}$ to decrease with cell density for the set of scratch assay data considered in this work. Concretely, the diffusion term $D_{\text{MLP}}$ was constrained to values between 0.0 and 0.096 $\nicefrac{\text{mm}^2}{\text{day}}$ and the growth term $G_{\text{MLP}}$ to values between $-0.48$ and 2.4 $\nicefrac{1}{\text{day}}$. The maximum and minimum diffusion values and maximum growth value were chosen based on values used in~\cite{simpson_2016}. The minimum growth value was chosen to be negative $20\%$ of the maximum growth value to allow $G_{\text{MLP}}$ to output negative values for cell densities near the carrying capacity if needed. The sigmoid output activation function for the delay term $T_{\text{MLP}}$ constrained its outputs to between 0 and 1. Derivative terms were used in $\mathcal{L}_{\text{Constr}}$ to constrain $D_{\text{MLP}}$ and $T_{\text{MLP}}$ to be non-decreasing and $G_{\text{MLP}}$ to be non-increasing. For ease of notation, let $\hat{u}_{i,j} \equiv u_{\text{MLP}}(x_i,t_j)$, $\hat{D}_{i,j} \equiv D_{\text{MLP}}(u_{\text{MLP}}(x_i,t_j))$, $\hat{G}_{i,j} \equiv G_{\text{MLP}}(u_{\text{MLP}}(x_i,t_j))$, and $\hat{T}_{i,j} \equiv T_{\text{MLP}}(t_j)$, then the constraint term can be written concretely as 
\begin{align}
\label{eq:concrete_constraint_loss}
    \mathcal{L}_{\text{Constr}} = \frac{1}{MN} \Bigg[ 
    & 
    \alpha_1 \sum_{\substack{i=1,j=1 \\ \hat{D}<0.0 \\ \hat{D}>0.096}}^{M,N} \left(\hat{D}_{i,j}\right)^2 + 
    \alpha_2 \sum_{\substack{i=1,j=1 \\ \partial \hat{D}/\partial \hat{u}<0}}^{M,N} \left(\frac{\partial \hat{D}_{i,j}}{\partial \hat{u}_{i,j}}\right)^2 + 
    \\ &
    \alpha_3 \sum_{\substack{i=1,j=1 \\ \hat{G}<-0.48 \\ \hat{G}>2.4}}^{M,N} \left(\hat{G}_{i,j}\right)^2 + 
    \alpha_4 \sum_{\substack{i=1,j=1 \\ \partial \hat{G}/\partial \hat{u}<0}}^{M,N} \left(\frac{\partial \hat{G}_{i,j}}{\partial \hat{u}_{i,j}}\right)^2 + 
    \alpha_5 \sum_{\substack{i=1,j=1 \\ \partial \hat{T}/\partial \hat{t}<0}}^{M,N} \left(\frac{\partial \hat{T}_{i,j}}{\partial \hat{u}_{i,j}}\right)^2 \Bigg]. \nonumber
\end{align}
Since the parameter networks and their derivatives occur at different scales with respect to each other and with respect to the error terms $\mathcal{L}_{\text{GLS}}$ and $\mathcal{L}_{\text{PDE}}$, each term of Eq~\eqref{eq:concrete_constraint_loss} is weighted by a factor $\alpha_i$. In particular, each constraint is weighted based on the input/output scaling factors of the corresponding neural network (see Network Design subsection). Concretely, the terms in Eq~\eqref{eq:concrete_constraint_loss} are weighted by $\alpha_1=\nicefrac{1}{0.096}\times10^{10}$, $\alpha_2=\nicefrac{K}{0.096}\times10^{10}$, $\alpha_3=\nicefrac{1}{2.4}\times10^{10}$, $\alpha_4=\nicefrac{K}{2.4}\times10^{10}$, and $\alpha_5=10^{10}$. Note that the weight factors for the derivative constraints on $D_{\text{MLP}}$ and $G_{\text{MLP}}$ (i.e. $\alpha_2$ and $\alpha_4$) include the carrying capacity $K=1.7\times10^3$ since $K$ was used as an input scaling factor for these networks. The factor $10^{10}$ was chosen large enough to guarantee that $D_{\text{MLP}}$, $G_{\text{MLP}}$, and $T_{\text{MLP}}$ exhibited the desired behavior. \JHL{Boundary conditions can also be included in the $\mathcal{L}_{\text{Constr}}$ term, however, since they were unknown for the scratch assay data considered in this work, no boundary conditions were used to train $u_{\text{MLP}}$.}

Finally, the GLS errors at the initial condition (i.e. data locations where $t=0$) were weighted by a factor of 10 during training. This was found to improve the generalization accuracy of $D_{\text{MLP}}$, $G_{\text{MLP}}$, and $T_{\text{MLP}}$ when evaluated using a numerical PDE solver. The reason for this is because the cell density at $t=0$ may not satisfy a governing dynamical system since the measurement is taken directly after the scratch assay protocol is performed~\cite{simpson_2016}. However, the initial condition ``sets the stage'' for the governing dynamics to drive the temporal evolution of the system. Therefore, by weighting the initial condition more heavily in $\mathcal{L}_{\text{GLS}}$, the PDE error term $\mathcal{L}_{\text{PDE}}$ must conform $u_{\text{MLP}}$ to satisfy the governing system for $t>0$ as dictated by $u_{\text{MLP}}$ at $t=0$. This step forced $D_{\text{MLP}}$, $G_{\text{MLP}}$, and $T_{\text{MLP}}$ to learn more generalizable representations of the diffusivity, growth, and delay functions, respectively. The weighting factor was numerically validated using the mean GLS error across each scratch assay experiment for weighting factors $1$, $10$, and $10^2$. \JHL{Note that this weighting factor makes BINNs sensitive to the random choice of training/validation split, since some data points in the initial condition may be more informative than others for equation learning and ultimate model generalizability. This observation was also noted in a recent equation learning study in which the random split of training and validation sets was found to influence the structure of the learned equation \cite{lagergren_2020}. Adopting a strategy similar to this previous study, BINNs were trained 20 times for each data set (using different random training/validation splits). The BINN for which the numerical simulations resulted in the smallest GLS error was saved as the best model.} 

\subsection*{PDE Forward Solver}

The numerical implementation details are provided for systems describing quantity of interest $u(x,t)$ that are governed by the following equation:
\begin{align}
    u_t &= \big(Q(u,u_x,t)\big)_x + F(u), \label{eq:MOL} 
    \\
    u(x,t_0) &= \phi(x), \nonumber 
    \\
    u_x(x_0,t) &= u_x(x_f,t) = 0, \nonumber
\end{align}
for $x\in[x_0,x_f]$, and $t\in[t_0,t_f]$. Note that the reaction-diffusion model in Eq~\eqref{eq:reaction_diffusion} is an example of Eq~\eqref{eq:MOL} where $Q(u,u_x,t) = \mathcal{D}(u,t)u_x$ and $F(u) = \mathcal{G}(u,t) u$. In Eq~\eqref{eq:MOL}, the initial condition is denoted by $\phi(x)$ and the boundary conditions are assumed to be no-flux boundary conditions. \KBF{Note that the no-flux condition represents a zero net flux boundary condition which does not preclude cells moving across the boundary, but instead reflects the situation in which the flux in the positive and negative $x$-directions are equal, giving rise to zero total flux.} The spatial and temporal domains are discretized into equispaced grids as:
\begin{eqnarray}
    x_i = i\Delta x, \quad  t_j = j\Delta t,
\end{eqnarray}
for $i = 0, \dots, 200$ and $j = 0, \dots, 1,000$. For notational convenience, let $u_i(t) = u(x_i,t)$. Then, the method-of-lines approach is used to solve Eq~\eqref{eq:MOL} with the numerical discretization from~\cite{kurganov_new_2000} that is given by
\begin{eqnarray}
     \big(Q(u,u_x,t)\big)_x \approx \dfrac{P_{i+1/2}(t) - P_{i-1/2}(t)}{\Delta x},
\end{eqnarray}
where $P_{i+1/2}(t)$ is an estimate for the rightwards diffusive flux at location $x_i$ that is given by
\begin{eqnarray}
    P_{i+1/2}(t) = \dfrac{1}{2}\left[ Q\left(u_i(t),\dfrac{u_{i+1}(t)-u_i(t)}{\Delta x},t \right) + Q\left(u_{i+1}(t),\dfrac{u_{i+1}(t)-u_i(t)}{\Delta x},t \right) \right].
\end{eqnarray}
The no-flux boundary conditions at $x_0$ and $x_{200}$ are implemented by incorporating the ghost points $x_{-1}$ and $x_{201}$ satisfying $u_{-1}(t) = u_1(t)$ and $u_{201}(t) = u_{199}(t)$. The Scipy integration subpackage (version 1.4.1) is used to integrate Eq~\eqref{eq:MOL} over time using an explicit fourth order Runge-Kutta Method. 

\subsection*{Parameter estimation}

The parameters of each mechanistic model were optimized using the Limited-memory BFGS algorithm with bound constraints (L-BFGS-B) in Python's Scipy package with default tolerance values to minimize the generalized least squares error function in Eq~\eqref{eq:gls_loss} with the adjusted statistical error model in Eq~\eqref{eq:statistical_model} with $\gamma=0.2$. The parameters for Eqs~\eqref{eq:classical_fkpp} and \eqref{eq:porous_fkpp} were initialized using the values from~\cite{simpson_2016}. The parameters for Eq~\eqref{eq:mechanistic_delay_reaction_diffusion} were initialized by fitting each PDE term in Eqs~\eqref{eq:mechanistic_diffusion}-\eqref{eq:mechanistic_delay} to the corresponding parameter network solutions in Fig~\ref{fig:delay_rd_binn_parameters} using ordinary least squares. Finally, the diffusivity and growth function parameters were bounded using $D_{\text{min}}=0$ $\nicefrac{\text{mm}^2}{\text{day}}$, $D_{\text{max}}=0.096$ $\nicefrac{\text{mm}^2}{\text{day}}$, $m_{\text{min}}=0$, $m_{\text{max}}=4$, $r_{\text{min}}=0$ $\nicefrac{1}{\text{day}}$, and $r_{\text{max}}=2.4$ $\nicefrac{1}{\text{day}}$ (all of which come from~\cite{simpson_2016}), while the delay function parameters $\beta_0$ and $\beta_1$ were bounded by $[-10, 10]$.

\section*{Acknowledgments}
This material was based upon work partially supported by the National Science Foundation under Grant DMS-1638521 to the Statistical and Applied Mathematical Sciences Institute and IOS-1838314 to KBF, and in part by National Institute of Aging grant R21AG059099 to KBF. Any opinions, findings, and conclusions or recommendations expressed in this material are those of the authors and do not necessarily reflect the views of the National Science Foundation. REB is a Royal Society Wolfson Research Merit Award holder and also acknowledges the Biotechnology and Biological Sciences Research Council for funding via grant no. BB/R000816/1. MJS acknowledges the Australian Research Council (DP200100177).

\nolinenumbers

% Either type in your references using
% \begin{thebibliography}{}
% \bibitem{}
% Text
% \end{thebibliography}
%
% or
%
% Compile your BiBTeX database using our plos2015.bst
% style file and paste the contents of your .bbl file
% here. See http://journals.plos.org/plosone/s/latex for 
% step-by-step instructions.
% 

% temporary
%\newpage
%\bibliography{references}

\newpage
\section*{Supporting information}

% Include only the SI item label in the paragraph heading. Use the~\nameref{label} command to cite SI items in the text.

\paragraph*{Fig S1}
\label{scratch_assay_experiment}
{\bf Scratch assay experiment.} \KBF{(a) An illustration of an experiment with the IncuCyte ZOOM\texttrademark\space system (Essen BioScience, MI USA). Full details of the experiment and image processing can be found in \cite{simpson_2016}. Cells are seeded uniformly within each well in a 96-well plate at a pre-specified density of between 10,000 and 20,000 cells per well. A WoundMaker\texttrademark\space (Essen BioScience) is used to create a uniform vertical scratch along the middle of the well. (b) Microscopy images are collected from a rectangular region of the well. (c) Example images corresponding to experiments initiated with 12,000, 16,000, or 20,000 cells per well. A PC-3 prostate cancer cell line was used. The image recording time is indicated on each subfigure and the scale bar corresponds to 300 $\mu$m. The green dashed lines in the images in the top row show the approximate location of the leading edge created by the scratch. Each image is divided into equally-spaced vertical columns, and the number of cells in each column divided by the column area is calculated to yield an estimate of the 1-D cell density.}
\begin{figure}[!h]
    \centering
    \includegraphics[width=1.0\textwidth]{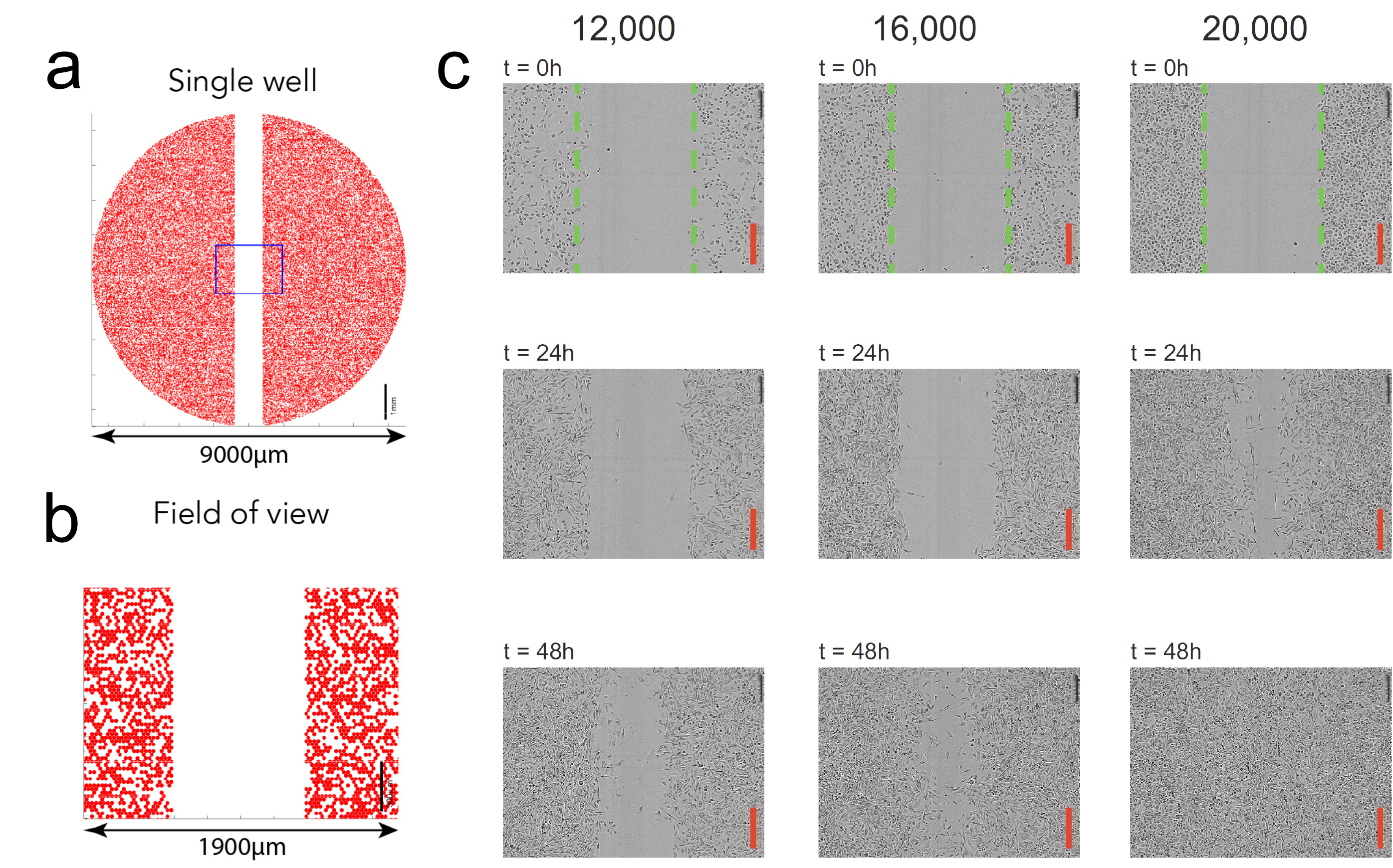}
\end{figure}

\newpage
\paragraph*{Fig S2}
\label{sim_model_fit}
{\bf Simulation model fits.} Predicted cell density profiles using BINNs with the governing reaction-diffusion PDE in Eq~\eqref{eq:reaction_diffusion_binn}. The left subplot corresponds to the set of simulated data using the classical FKPP equation and the right subplot corresponds to the Generalized Porous-FKPP equation. Solid lines represent the numerical solution to Eq~\eqref{eq:reaction_diffusion_binn} using $D_{\text{MLP}}$, and $G_{\text{MLP}}$. Dashed lines represent the noiseless numerical simulations of the classical FKPP and Generalized Porous-FKPP equations. The markers represent the numerical simulations of the classical FKPP and Generalized Porous-FKPP equations with artificial noise generated by the statistical error model in Eq~\eqref{eq:statistical_model}.
\begin{figure}[!h]
    \centering
    \includegraphics[width=1.0\textwidth]{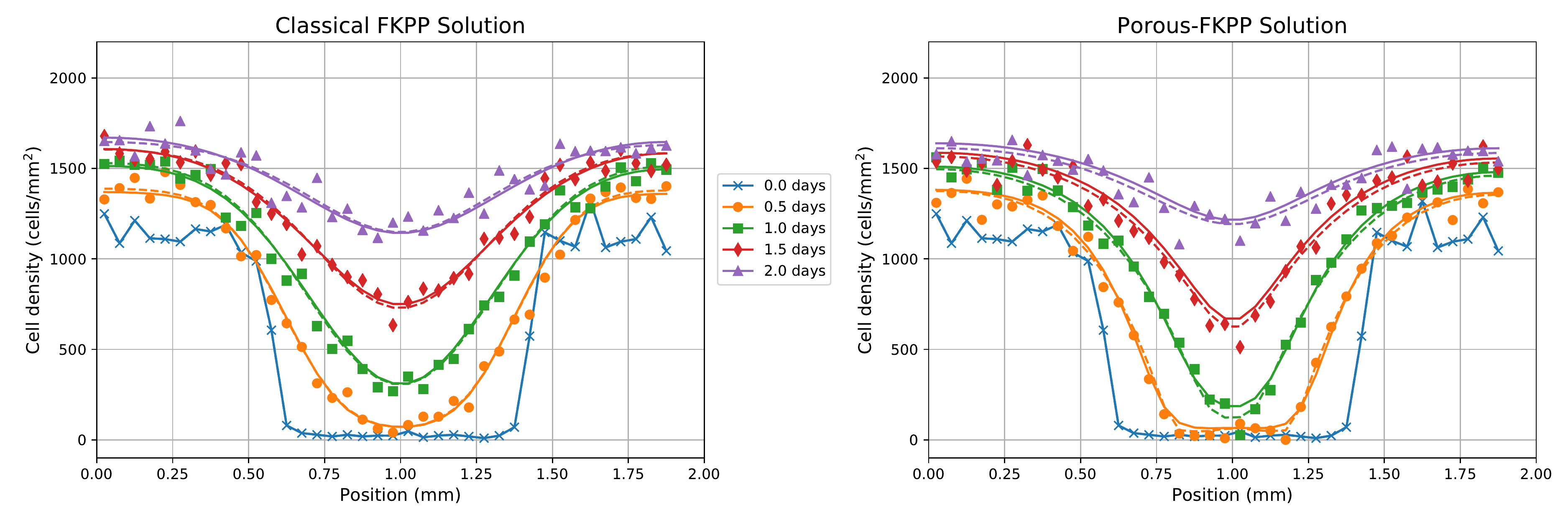}
\end{figure}

\paragraph*{Fig S3}
\label{sim_parameter_fit}
{\bf Simulation parameter fits.} The learned diffusivity and growth functions $D_{\text{MLP}}$ and $G_{\text{MLP}}$ evaluated over cell density $u$. Starting from the left, the first two subplots correspond to the learned diffusivity and growth functions from simulated data using the classical FKPP equation. The last two subplots correspond to the learned diffusivity and growth functions from simulated data using the Generalized Porous-FKPP equation. Solid lines represent the parameter networks $D_{\text{MLP}}$ and $G_{\text{MLP}}$ and dashed lines represent the true diffusivity and growth functions used to simulate the data.
\begin{figure}[!h]
    \centering
    \includegraphics[width=1.0\textwidth]{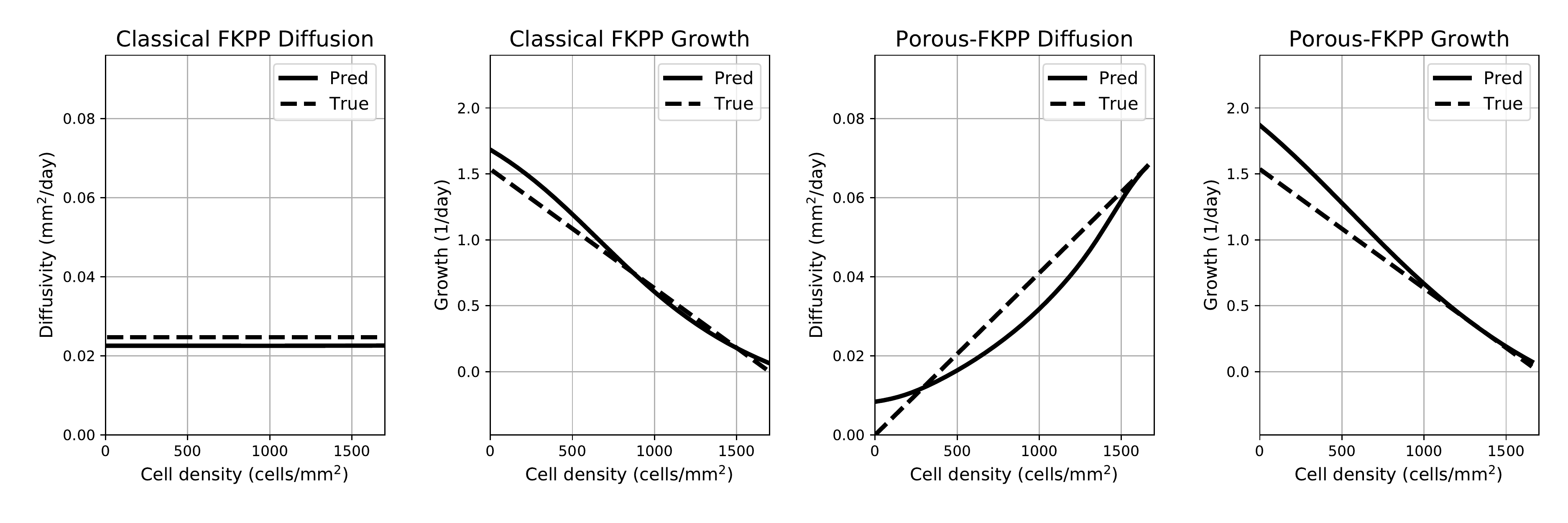}
\end{figure}

\newpage
\paragraph*{Fig S4}
\label{rdbinn_residuals}
\textbf{Reaction-diffusion BINN residuals.} Modified residuals using BINNs with the governing reaction-diffusion PDE in Eq~\eqref{eq:reaction_diffusion_binn}. Each subplot corresponds to an experiment with a different initial cell density (i.e. 10,000, 12,000, 14,000, 16,000, 18,000, and 20,000 cells per well). 
\begin{figure}[!h]
    \centering
    \includegraphics[width=1.0\textwidth]{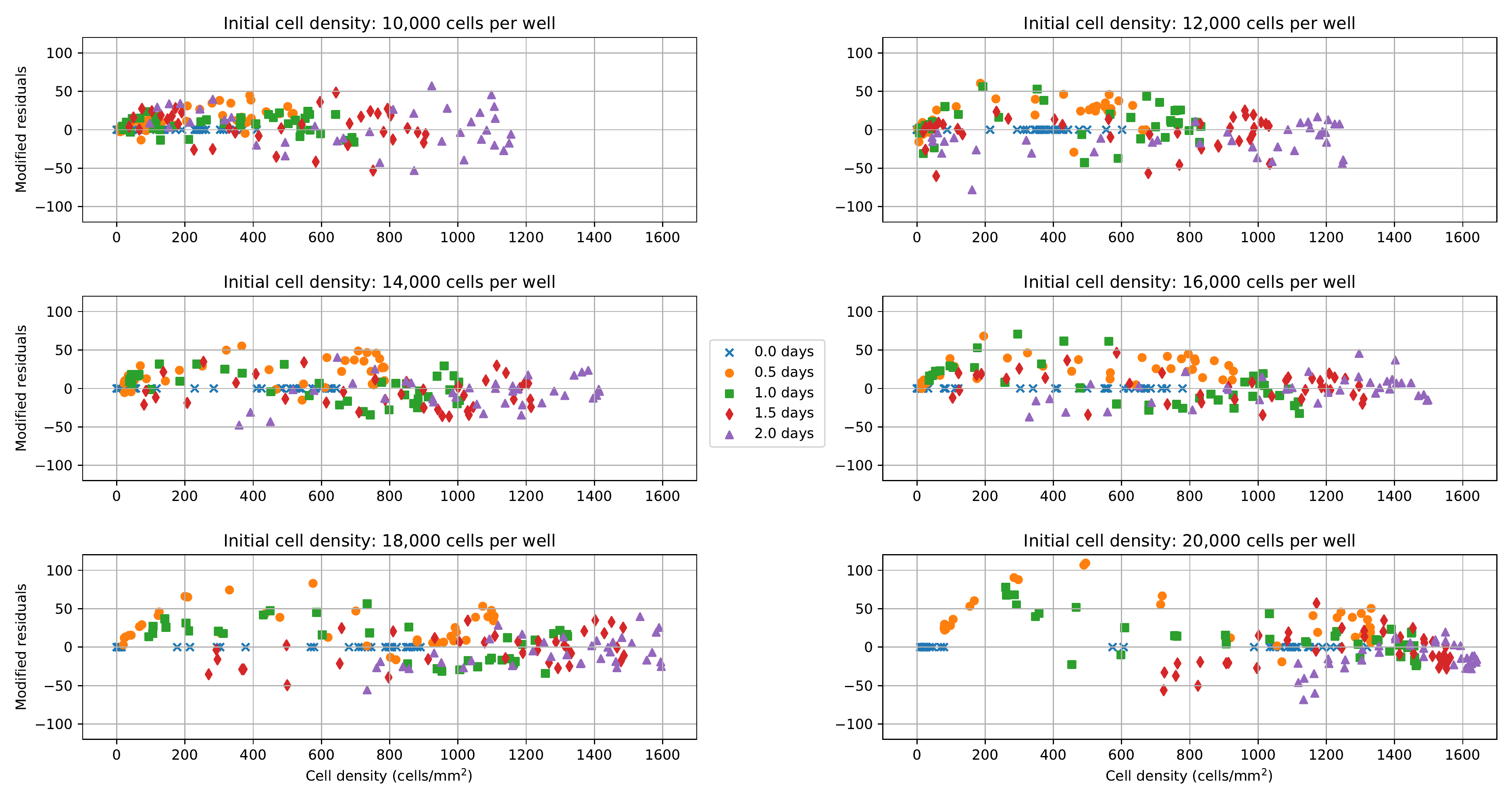}
\end{figure}

\paragraph*{Fig S5}
\label{drdbinn_residuals}
\textbf{Delay-reaction-diffusion BINN residuals.} Modified residuals using BINNs with the governing delay-reaction-diffusion PDE in Eq~\eqref{eq:delay_reaction_diffusion_binn}. Each subplot corresponds to an experiment with a different initial cell density (i.e. 10,000, 12,000, 14,000, 16,000, 18,000, and 20,000 cells per well). 
\begin{figure}[!h]
    \centering
    \includegraphics[width=1.0\textwidth]{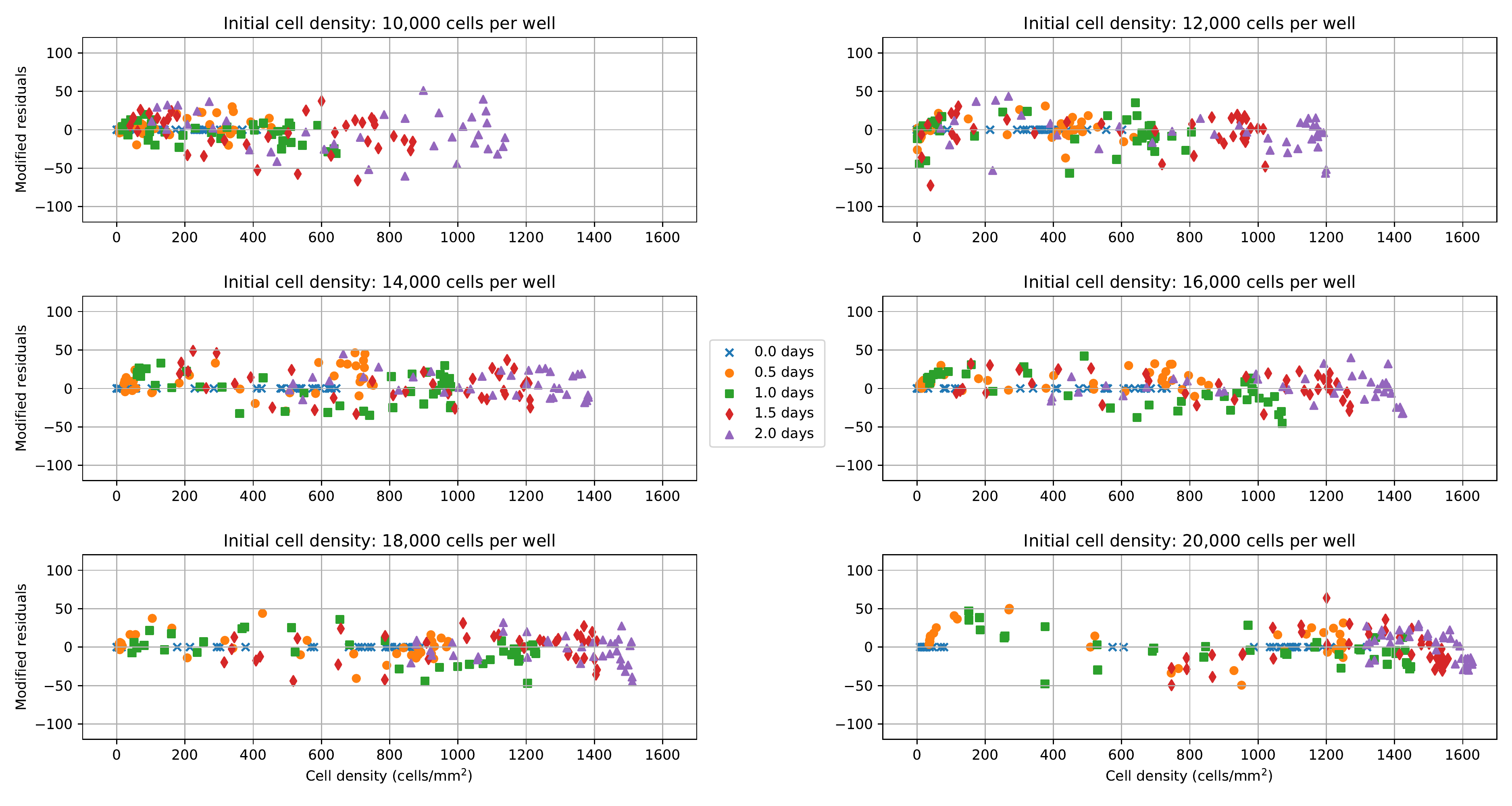}
\end{figure}

\newpage
\paragraph*{Fig S6}
\label{fkpp_solutions}
\textbf{Classical FKPP model solutions.} Predicted cell density profiles using the classical FKPP model in Eq~\eqref{eq:classical_fkpp}. Each subplot corresponds to an experiment with a different initial cell density (i.e. 10,000, 12,000, 14,000, 16,000, 18,000, and 20,000 cells per well). Solid lines represent the numerical solution to Eq~\eqref{eq:classical_fkpp} using the parameters that minimize $\mathcal{L}_{\text{GLS}}$ in Eq~\eqref{eq:gls_loss}. The markers represent the experimental scratch assay data.
\begin{figure}[!h]
    \centering
    \includegraphics[width=1.0\textwidth]{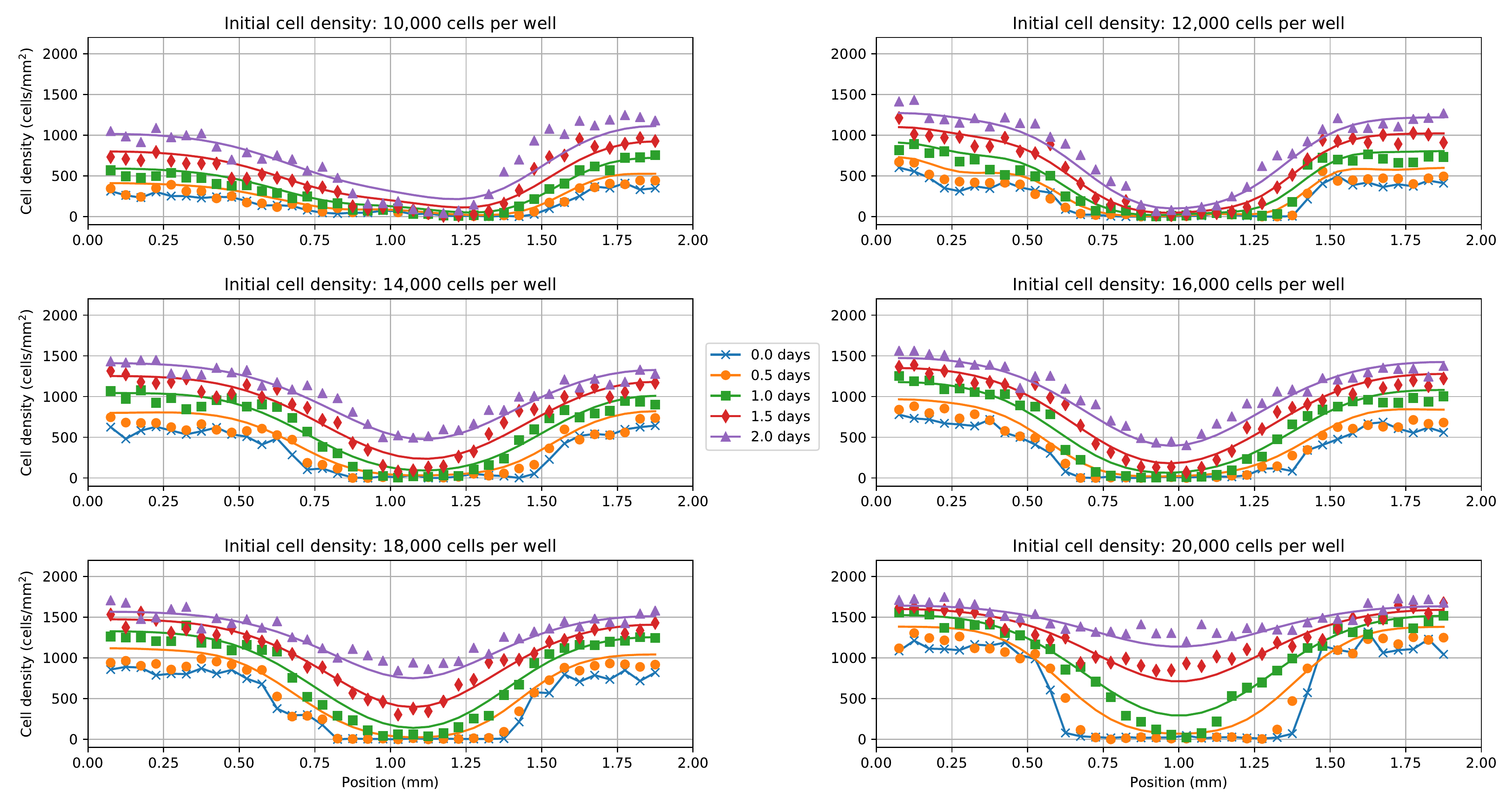}
\end{figure}

\paragraph*{Fig S7}
\label{pfkpp_solutions}
\textbf{Generalized Porous-FKPP model solutions.} Predicted cell density profiles using the Generalized Porous-FKPP model in Eq~\eqref{eq:porous_fkpp}. Each subplot corresponds to an experiment with a different initial cell density (i.e. 10,000, 12,000, 14,000, 16,000, 18,000, and 20,000 cells per well). Solid lines represent the numerical solution to Eq~\eqref{eq:porous_fkpp} using the parameters that minimize $\mathcal{L}_{\text{GLS}}$ in Eq~\eqref{eq:gls_loss}. The markers represent the experimental scratch assay data.
\begin{figure}[!h]
    \centering
    \includegraphics[width=1.0\textwidth]{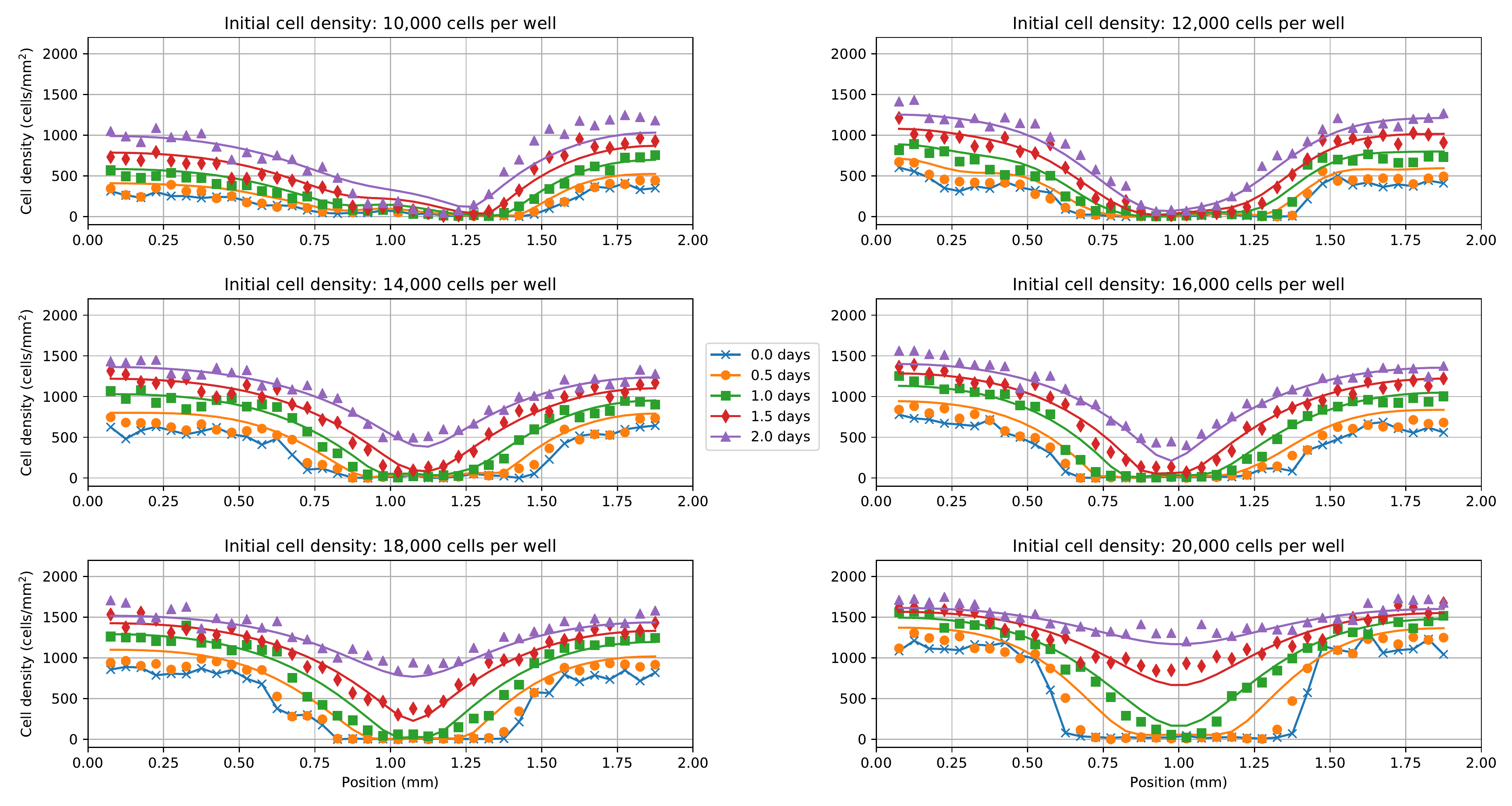}
\end{figure}

\newpage
\paragraph*{Tab S1}
\label{fkpp_parameters}
\textbf{Classical FKPP parameter values.} Each column corresponds to an experiment with different initial cell density (i.e. 10,000, 12,000, 14,000, 16,000, 18,000, and 20,000 cells per well).
\begin{table}[!h]
    \centering
    \begin{tabular}{l|cccccc}
        \multicolumn{1}{l}{} & \multicolumn{6}{c}{Initial cell density} \\
        Parameter & 10,000 & 12,000 & 14,000 & 16,000 & 18,000 & 20,000
        \\ \hline
        $D$ ($\nicefrac{\mu\text{m}^2}{\text{hr}}$) & 309.7 & 253.8 & 681.8 & 540.9 & 735.7 & 978.5 \\
        $r$ ($\nicefrac{1}{\text{hr}}$) & 0.0437 & 0.0438 & 0.0483 & 0.0490 & 0.0540 & 0.0649 \\
    \end{tabular}
\end{table}

%$D$ ($\nicefrac{\text{mm}^2}{\text{day}}$) & 0.0074 & 0.0061 & 0.0164 & 0.0130 & 0.0177 & 0.0235 \\
%$r$ ($\nicefrac{1}{\text{day}}$) & 1.0476 & 1.0510 & 1.1586 & 1.1772 & 1.2961 & 1.5587 

\paragraph*{Tab S2}
\label{pfkpp_parameters}
\textbf{Generalized Porous-FKPP parameter values.} Each column corresponds to an experiment with different initial cell density (i.e. 10,000, 12,000, 14,000, 16,000, 18,000, and 20,000 cells per well).
\begin{table}[!h]
    \centering
    \begin{tabular}{l|cccccc}
        \multicolumn{1}{l}{} & \multicolumn{6}{c}{Initial cell density} \\
        Parameter & 10,000 & 12,000 & 14,000 & 16,000 & 18,000 & 20,000
        \\ \hline
        $D$ ($\nicefrac{\mu\text{m}^2}{\text{hr}}$) & 1851.8 &  465.0 & 2993.4 & 2371.0 & 2377.8 & 2017.1 \\
        $m$ (unitless) & 0.9704 & 0.3001 & 0.9923 & 0.9879 & 0.8265 & 0.6235 \\
        $r$ ($\nicefrac{1}{\text{hr}}$) & 0.0435 & 0.0436 & 0.0481 & 0.0484 & 0.0526 & 0.0639 \\
    \end{tabular}
\end{table}

%$D$ ($\nicefrac{\mu\text{m}^2}{\text{hr}}$) & 0.0444 & 0.0112 & 0.0718 & 0.0569 & 0.0571 & 0.0484 \\
%$m$ (unitless) & 0.9704 & 0.3001 & 0.9923 & 0.9879 & 0.8265 & 0.6235 \\
%$r$ ($\nicefrac{1}{\text{hr}}$) & 1.0429 & 1.0461 & 1.1541 & 1.1618 & 1.2621 & 1.5341 

\paragraph*{Fig S8}
\label{unconstr_binn_parameters}
\textbf{Unconstrained BINN terms.} The learned diffusivity $D_{\text{MLP}}$, growth $G_{\text{MLP}}$, and delay $T_{\text{MLP}}$ functions extracted from the corresponding BINNs with governing reaction-diffusion PDE in Eq~\eqref{eq:reaction_diffusion_binn} (first row) and delay-reaction-diffusion PDE in Eq~\eqref{eq:delay_reaction_diffusion_binn} (second row). Each line corresponds to an experiment with a different initial cell density (i.e. 10,000, 12,000, 14,000, 16,000, 18,000, and 20,000 cells per well). Note that $D_{\text{MLP}}$ and $G_{\text{MLP}}$ have different lengths since they are evaluated between the minimum and maximum observed cell densities corresponding to each data set.
\begin{figure}[!h]
    \centering
    \includegraphics[width=\columnwidth]{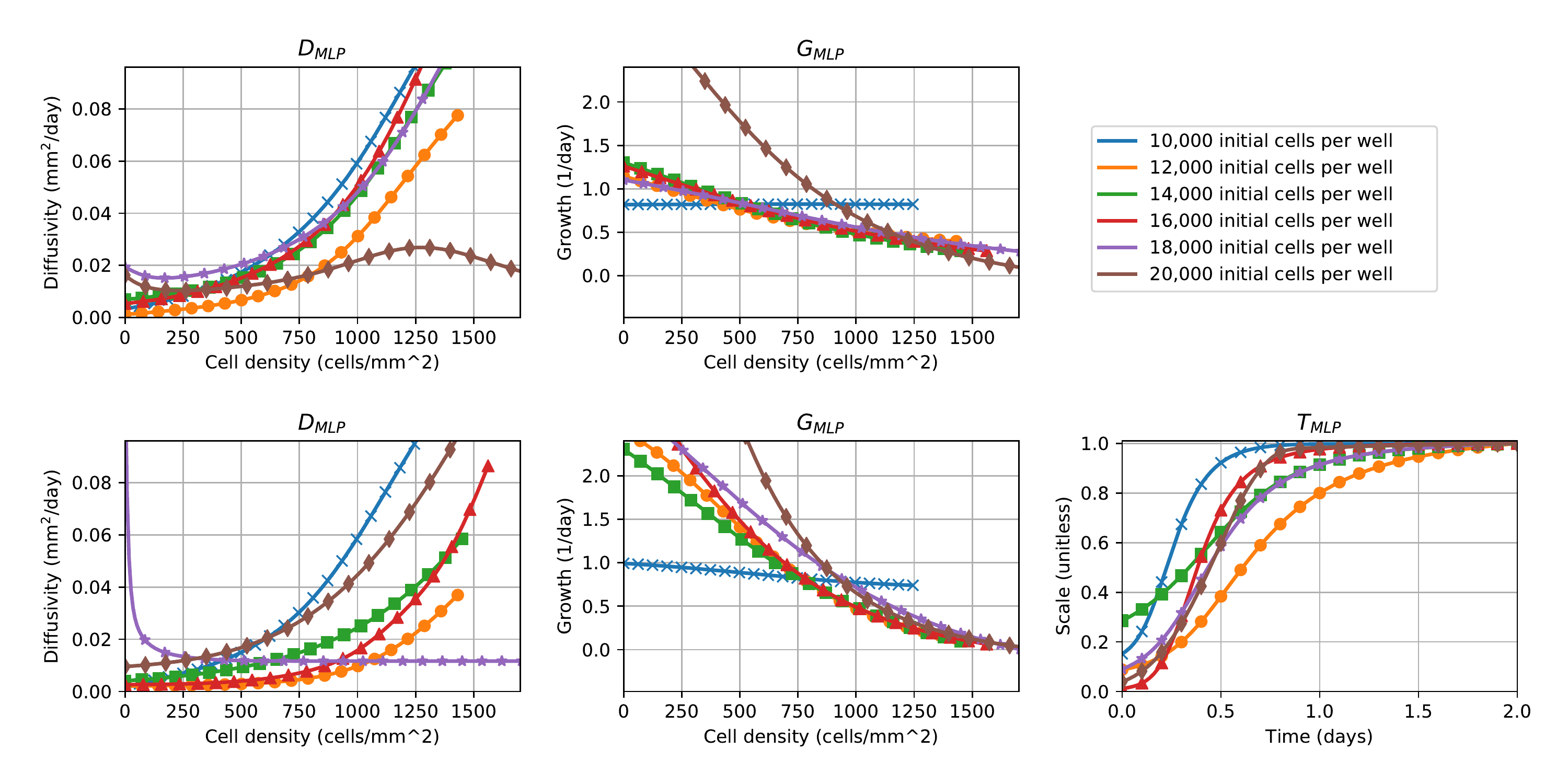}
\end{figure}

\newpage
\paragraph*{Fig S9}
\label{stat_model_selection}
{\bf Statistical error model selection.} The function-approximating deep neural network $u_{\text{MLP}}$ is trained using $\mathcal{L}_{\text{GLS}}$ for different values of $\gamma$ across each data set. Each subplot shows the modified residuals (see Eq~\eqref{eq:gls_loss}) as a function of the predicted cell density $u$. The columns correspond to different levels of proportionality (i.e. $\gamma=0.0, 0.2, 0.4, 0.6$) where $\gamma=0.0$ represents the constant variance (ordinary least squares) case. Each row (a-f) corresponds to an experiment with different initial cell density (i.e. 10,000, 12,000, 14,000, 16,000, 18,000, and 20,000 cells per well). The proportionality constant that results in the most i.i.d. residuals across each data set was chosen to calibrate the statistical error model in Eq~\eqref{eq:statistical_model}.
\begin{figure}[!h]
    \centering
    \includegraphics[width=\columnwidth]{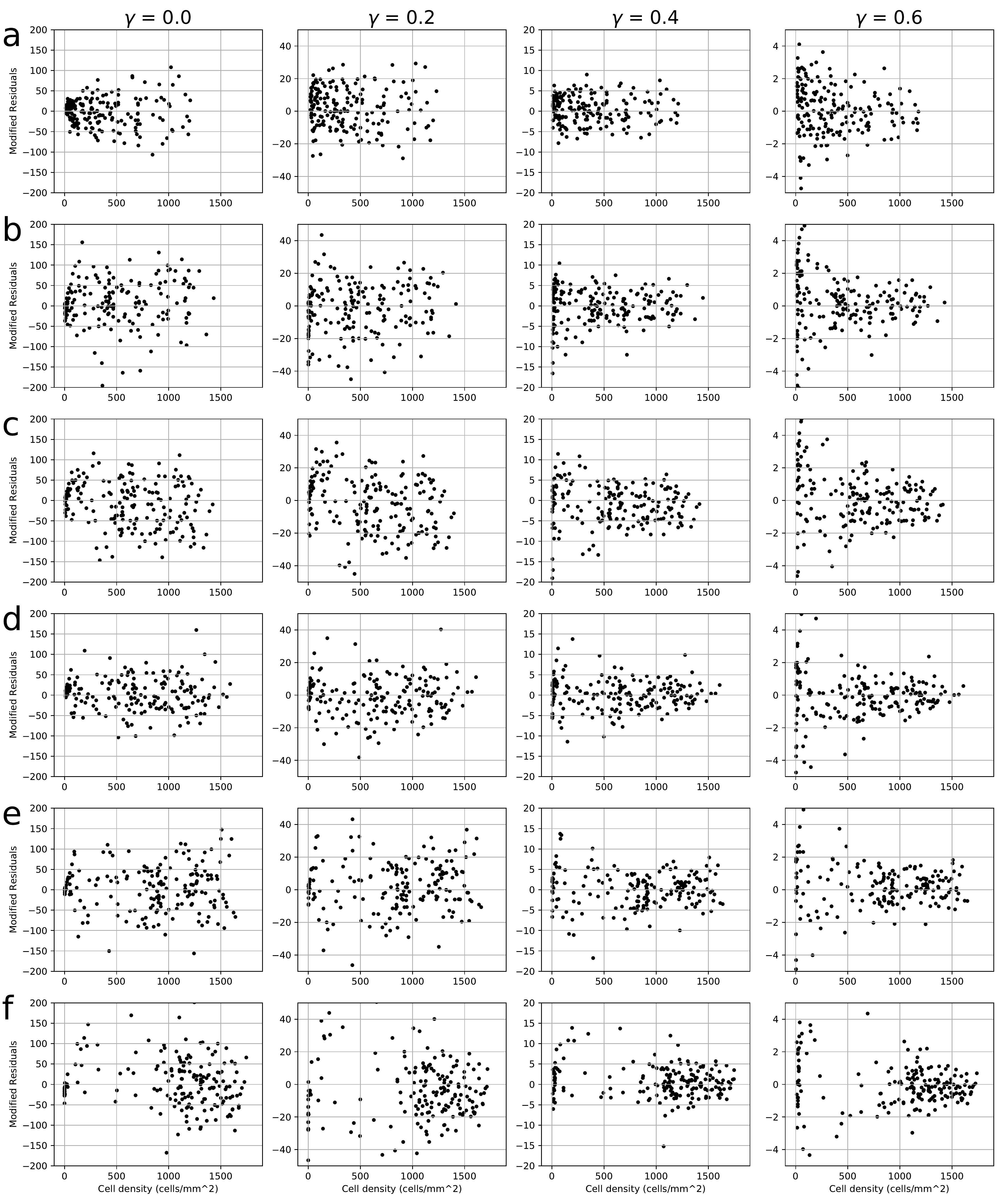}
\end{figure}

\newpage
\paragraph*{Fig S10}
\label{pde_random_sampling}
{\bf PDE random sampling validation.} The BINNs framework is trained using $\mathcal{L}_{\text{Total}}$ with three ways of including the PDE error term $\mathcal{L}_{\text{Total}}$: (a) no PDE regularization, (b) PDE regularization at the data locations, and (c) PDE regularization at 10,000 randomly sampled points at each training iteration. The first column shows the scratch assay data with initial cell density 20,000 cells per well (black dots) with the corresponding BINNs approximation to the governing PDE $u_{\text{MLP}}$ (surface plot). The second column shows heatmaps of the modified residual errors (see Eq~\eqref{eq:gls_loss}) at each data point. The third column shows heatmaps of the PDE errors (see Eq~\eqref{eq:pde_loss}) evaluated on a $100\times100$ meshgrid over the input domain. 
\begin{figure}[!h]
    \centering
    \includegraphics[width=\columnwidth]{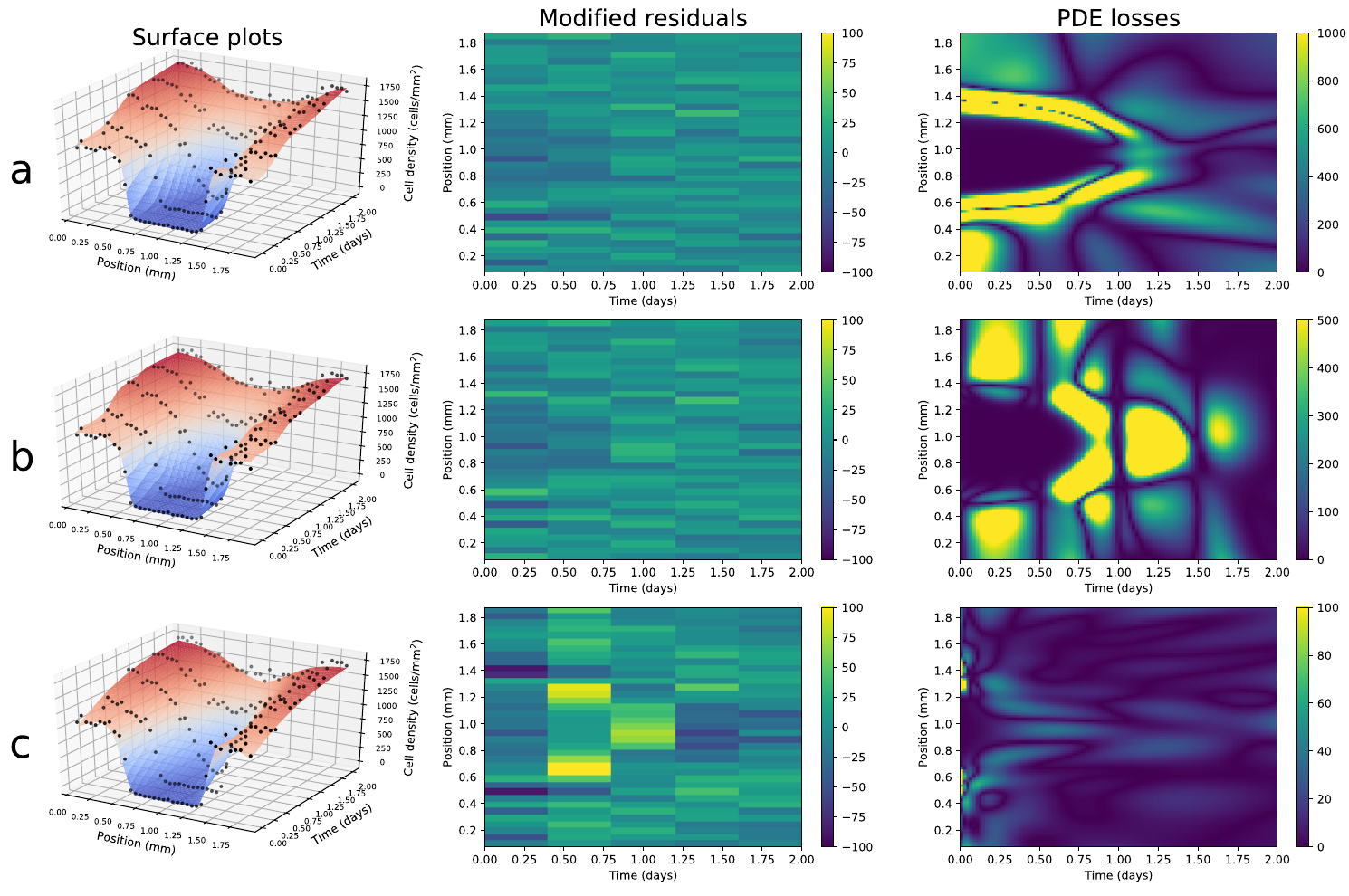}
\end{figure}

\end{document}